\documentclass[review]{elsarticle}

\usepackage{lineno,hyperref}
\modulolinenumbers[5]

\journal{Nuclear Instruments and Methods A}

\begin{document}

\begin{frontmatter}

\title{MINERvA neutrino detector response measured with test beam data}


\newcommand{\Rutgers}{Rutgers, The State University of New Jersey, Piscataway, New Jersey 08854, USA}
\newcommand{\Hampton}{Hampton University, Dept. of Physics, Hampton, VA 23668, USA}
\newcommand{\Dortmund}{Institute of Physics, Dortmund University, 44221, Germany }
\newcommand{\Otterbein}{Department of Physics, Otterbein University, 1 South Grove Street, Westerville, OH, 43081 USA}
\newcommand{\JMU}{James Madison University, Harrisonburg, Virginia 22807, USA}
\newcommand{\Florida}{University of Florida, Department of Physics, Gainesville, FL 32611}
\newcommand{\UCIrvine}{Department of Physics and Astronomy, University of California, Irvine, Irvine, California 92697-4575, USA}
\newcommand{\CBPF}{Centro Brasileiro de Pesquisas F\'{i}sicas, Rua Dr. Xavier Sigaud 150, Urca, Rio de Janeiro, Rio de Janeiro, 22290-180, Brazil}
\newcommand{\PUCP}{Secci\'{o}n F\'{i}sica, Departamento de Ciencias, Pontificia Universidad Cat\'{o}lica del Per\'{u}, Apartado 1761, Lima, Per\'{u}}
\newcommand{\INRM}{Institute for Nuclear Research of the Russian Academy of Sciences, 117312 Moscow, Russia}
\newcommand{\Jlab}{Jefferson Lab, 12000 Jefferson Avenue, Newport News, VA 23606, USA}
\newcommand{\Pittsburgh}{Department of Physics and Astronomy, University of Pittsburgh, Pittsburgh, Pennsylvania 15260, USA}
\newcommand{\Guanajuato}{Campus Le\'{o}n y Campus Guanajuato, Universidad de Guanajuato, Lascurain de Retana No. 5, Colonia Centro, Guanajuato 36000, Guanajuato M\'{e}xico.}
\newcommand{\Athens}{Department of Physics, University of Athens, GR-15771 Athens, Greece}
\newcommand{\Tufts}{Physics Department, Tufts University, Medford, Massachusetts 02155, USA}
\newcommand{\WM}{Department of Physics, College of William \& Mary, Williamsburg, Virginia 23187, USA}
\newcommand{\FNAL}{Fermi National Accelerator Laboratory, Batavia, Illinois 60510, USA}
\newcommand{\Purdue}{Department of Chemistry and Physics, Purdue University Calumet, Hammond, Indiana 46323, USA}
\newcommand{\MCLA}{Massachusetts College of Liberal Arts, 375 Church Street, North Adams, MA 01247}
\newcommand{\UMD}{Department of Physics, University of Minnesota -- Duluth, Duluth, Minnesota 55812, USA}
\newcommand{\Northwestern}{Northwestern University, Evanston, Illinois 60208}
\newcommand{\UNI}{Universidad Nacional de Ingenier\'{i}a, Apartado 31139, Lima, Per\'{u}}
\newcommand{\Rochester}{University of Rochester, Rochester, New York 14627 USA}
\newcommand{\Austin}{Department of Physics, University of Texas, 1 University Station, Austin, Texas 78712, USA}
\newcommand{\USM}{Departamento de F\'{i}sica, Universidad T\'{e}cnica Federico Santa Mar\'{i}a, Avenida Espa\~{n}a 1680 Casilla 110-V, Valpara\'{i}so, Chile}
\newcommand{\Geneva}{University of Geneva, 1211 Geneva 4, Switzerland}
\newcommand{\Chicago}{Enrico Fermi Institute, University of Chicago, Chicago, IL 60637 USA}
\newcommand{\hired}{}
\newcommand{\bmeThanks}{now at SLAC National Accelerator Laboratory, Stanford, California 94309 USA}
\newcommand{\higueraThanks}{University of Houston, Houston, Texas, 77204, USA}
\newcommand{\LazaThanks}{also at Department of Physics, University of Antananarivo, Madagascar}
\newcommand{\ticeThanks}{now at Argonne National Laboratory, Argonne, IL 60439, USA }
\newcommand{\twaltonThanks}{now at Fermi National Accelerator Laboratory, Batavia, IL USA 60510}

\author[WM, PUCP]     {L.~Aliaga}
\author[Tufts]        {O.~Altinok}
\author[PUCP]   {C.~Araujo~Del~Castillo}
\author[FNAL]         {L.~Bagby}
\author[FNAL]         {L.~Bellantoni}
\author[WM]      {W.F.~Bergan}
\author[Rochester]    {A.~Bodek}
\author[Rochester]    {R.~Bradford\thanks{\ticeThanks}}
\author[Geneva]    {A.~Bravar}
\author[Rochester]    {H.~Budd}
\author[INRM]         {A.~Butkevich}
\author[CBPF,FNAL]    {D.A.~Martinez~Caicedo}
\author[CBPF]         {M.F.~Carneiro}
\author[Hampton]      {M.E.~Christy}
\author[Rochester]    {J.~Chvojka}
\author[CBPF]         {H.~da~Motta}
\author[WM]           {J.~Devan}
\author[Rochester,PUCP]{G.A.~D\'{i}az~}
\author[Pittsburgh]   {S.A.~Dytman}
\author[Pittsburgh]   {B.~Eberly\thanks{\bmeThanks}}
\author[Guanajuato]   {J.~Felix}
\author[Northwestern] {L.~Fields}
\author[Rochester]    {R.~Fine}
\author[Rochester]    {R.~Flight}
\author[PUCP]         {A.M.~Gago}
\author[FNAL]         {C.~Gingu}
\author[Rochester,FNAL]{T.~Golan}
\author[Rochester]    {A.~Gomez}
\author[UMD]          {R.~Gran}
\author[FNAL]         {D.A.~Harris}
\author[Rochester,Guanajuato]{A.~Higuera\thanks{\higueraThanks}}
\author[WM]      {I.J.~Howley}
\author[CBPF,UNI]     {K.~Hurtado}
\author[Rochester]    {J.~Kleykamp}
\author[WM]           {M.~Kordosky}
\author[UMD]      {M.~Lanari}
\author[Rutgers]      {T.~Le}
\author[WM]      {A.J.~Leister}
\author[UMD]      {A.~Lovlein}
\author[MCLA]         {E.~Maher}
\author[Tufts]        {W.A.~Mann}
\author[Rochester]    {C.M.~Marshall}
\author[Rochester,FNAL]{K.S.~McFarland}
\author[Pittsburgh]   {C.L.~McGivern}
\author[Rochester]    {A.M.~McGowan}
\author[Pittsburgh]   {B.~Messerly}
\author[USM]          {J.~Miller}
\author[UMD]      {W.~Miller}
\author[Rochester]    {A.~Mislivec}
\author[FNAL]         {J.G.~Morf\'{i}n}
\author[Florida]      {J.~Mousseau}
\author[CBPF]         {T.~Muhlbeier}
\author[Pittsburgh]   {D.~Naples}
\author[WM]           {J.K.~Nelson}
\author[WM]           {A.~Norrick}
\author[PUCP]         {N.~Ochoa}
\author[WM]      {C.D.~O'Connor}
\author[Florida]      {B.~Osmanov}
\author[FNAL]         {J.~Osta}
\author[Pittsburgh]   {V.~Paolone}
\author[Northwestern] {C.E.~Patrick}
\author[Northwestern] {L.~Patrick}
\author[FNAL,Rochester] {G.N.~Perdue}
\author[PUCP]    {C.E.~P\'{e}rez~Lara}
\author[FNAL]         {L.~Rakotondravohitra\thanks{\LazaThanks}}
\author[Florida]      {H.~Ray}
\author[Pittsburgh]   {L.~Ren}
\author[Rochester]    {P.A.~Rodrigues}
\author[FNAL]  {P.~Rubinov}
\author[UMD]       {C.R.~Rude}
\author[Rochester]    {D.~Ruterbories}
\author[Northwestern] {H.~Schellman}
\author[Chicago,FNAL] {D.W.~Schmitz}
\author[UNI]          {C.J.~Solano~Salinas}
\author[Otterbein]    {N.~Tagg}
\author[Rutgers]      {B.G.~Tice\thanks{\ticeThanks}}
\author[Guanajuato]  {Z.~Urrutia}
\author[Guanajuato]   {E.~Valencia}
\author[Hampton]      {T.~Walton\thanks{\twaltonThanks}}
\author[UMD]      {A.~Westerberg}
\author[Rochester]    {J.~Wolcott}
\author[UMD]      {N.~Woodward}
\author[Florida]      {M.~Wospakrik}
\author[Guanajuato]   {G.~Zavala}
\author[WM]           {D.~Zhang}
\author[UCIrvine]     {B.P.~Ziemer}
\address[WM]{\WM}
\address[PUCP]{\PUCP}
\address[Tufts]{\Tufts}
\address[FNAL]{\FNAL}
\address[Rochester]{\Rochester}
\address[Geneva]{\Geneva}
\address[INRM]{\INRM}
\address[CBPF]{\CBPF}
\address[Hampton]{\Hampton}
\address[Pittsburgh]{\Pittsburgh}
\address[Guanajuato]{\Guanajuato}
\address[Northwestern]{\Northwestern}
\address[UMD]{\UMD}
\address[UNI]{\UNI}
\address[Rutgers]{\Rutgers}
\address[MCLA]{\MCLA}
\address[USM]{\USM}
\address[Florida]{\Florida}
\address[Otterbein]{\Otterbein}
\address[UCIrvine]{\UCIrvine}

\begin{abstract}
The MINERvA collaboration operated a scaled-down replica of the
solid scintillator tracking and sampling calorimeter regions of the
MINERvA detector in a hadron
test beam at the Fermilab Test Beam Facility.  
This article reports measurements with samples of protons, pions,
and electrons from 0.35 to 2.0~GeV/c momentum.  
The calorimetric response to protons,
pions, and electrons are obtained from these data.   
A measurement of the parameter in Birks' law and an estimate of the
tracking efficiency are extracted from the proton sample.
Overall the data are
well described by a Geant4-based Monte Carlo simulation of the
detector and particle interactions with
agreements better than 4\% for the calorimetric response, though some features of the data
are not precisely modeled.  These measurements are used
to tune the MINERvA detector simulation and evaluate systematic uncertainties in
support of the MINERvA neutrino cross section measurement program.
\bigskip
\\Fermilab preprint FERMILAB-PUB-15-018-ND
\end{abstract}


\end{frontmatter}


\section{Introduction and test beam goals}

The MINERvA experiment\cite{Aliaga:2013uqz} is designed to make
precision measurements of neutrino-nucleus cross sections.
An important part of 
these \cite{Fields:2013zhk,Fiorentini:2013ezn,Tice:2014pgu,Eberly:2014mra,Walton:2014esl}
and future cross section measurements is the estimate of the energy of one
or more hadrons exiting the nucleus.  These hadrons include recoil
protons and neutrons
with kinetic energies from hundreds of MeV to a few GeV,
pions from inelastic production, and softer nucleons
and nuclear fragments.  The goal of the test beam experiment is to
measure how well the Monte Carlo simulation (MC) of the detector response of these
particles describes the data.
The accuracy of the simulated single-particle response is an essential 
ingredient to the MINERvA neutrino cross section measurements.
Results presented in this paper include a
measurement of the Birks' law parameter \cite{Birks:1951,Birks:1964zz}, 
constraints on the accuracy of proton, pion, and electron calorimetry,
and a study of tracking efficiency for protons.

The detector used to take these data is a miniature replica of
MINERvA.  It is one meter in the transverse dimension, about half the size as MINERvA,
and one-third the depth.
These test beam data are the first from a new hadron beamline at the
Fermilab Test Beam Facility (FTBF) built for a data run in summer 2010
as Fermilab Test Beam Experiment T977.
There are differences between the two detectors that mitigate special aspects of the beam
environment in FTBF.  They include every-other-side readout and higher
light yield per MeV, and allow for a data set better suited for the
Birks' parameter and
calorimetry measurements.  

The energy range covered by these data, 0.35 to 2.0 GeV, is well-matched to the energy
range of protons, pions, and electromagnetic showers in the 2010 to 2012 MINERvA low-energy
neutrino and antineutrino data.  This is especially
true for the reactions from neutrino quasielastic scattering through
$\Delta$ and N$^*$ resonance production.   
Measuring differential cross sections for these
exclusive final states is a pillar of the MINERvA
neutrino physics program.
These energies also cover the lower part of the range expected for
hadrons produced in neutrino deep inelastic scattering.

This paper starts with a description of the test beam and associated instrumentation,
then the detector, followed by the resulting data sample with its simulation
and calibrations.  The Birks' law parameter measurement is next
because the parameter is used for all other
analyses.  Proton calorimetry results are presented followed by a
section with a complete discussion of systematic uncertainties for
proton, pion, and electron
measurements, which share the same sources but take on different
values.  With the  uncertainty discussion as a prelude, the pion calorimetry
results are described, followed by the the electron calorimetry
results, and then a discussion of calorimetry with respect to other
experimental results.  
The paper concludes with a measurement of tracking efficiency and a summary.

\section{Fermilab Test Beam Facility tertiary hadron beam}
\label{hadronbeam}

\begin{figure}[ht!]
\begin{center}
\includegraphics[width=12cm]{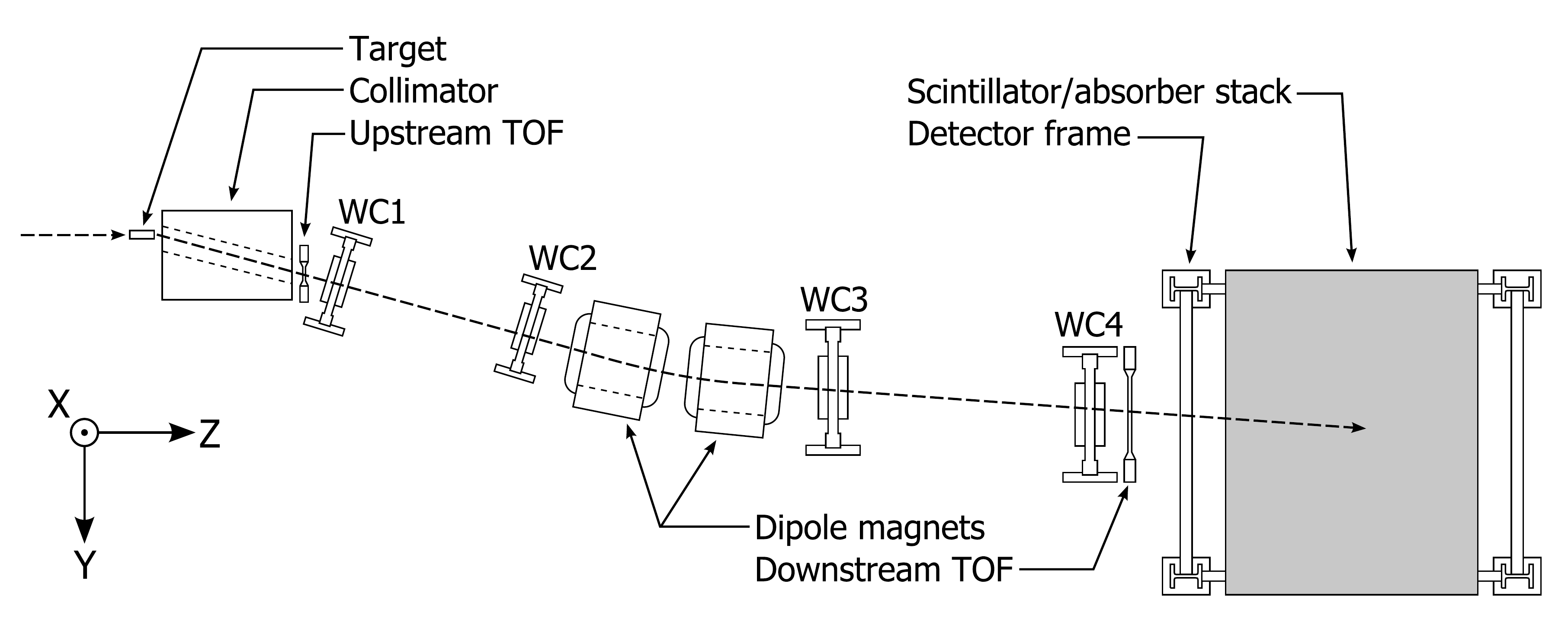}
\caption{ Diagram of the beamline built for this experiment, viewed
  from above with the beam going from left to right. }
\label{fig:beamlinediagram}
\end{center}
\end{figure}

This beam was developed through a partnership between the MINERvA
experiment and the Fermilab Test Beam Facility. 
It is produced from 16~GeV pions colliding with a copper
target. 
All species exit a collimator with an angle of 16 $\pm$ 1 degrees from
the direction of the incident pions.  
The species and momentum are tagged particle-by-particle using
time-of-flight (TOF) and position measurements from four wire
chambers.
Figure~\ref{fig:beamlinediagram} shows the geometry and
coordinate system viewed from the top with the beam propagating left
to right.  The incident 16~GeV pions initially encounter the target
and collimator, and the products of interactions in the target
continue toward two magnets.
The magnets are type ``NDB'' made at Fermilab, 
are ramped to a current of 100~A producing a 0.339 Tesla field in the
central region of the magnet, and have a polarity that can be reversed.
The typical field integral is 38.3~Tesla~cm with 1.5~Tesla~cm
variations around this value that encompass 90\% of selected events.
The detector, which is on the right, sees a range
of incident particles with low momentum at low horizontal (Y) coordinate and
normal incidence, with high momentum at higher Y-coordinates and
angles as far as 10 degrees from the detector axis.  In addition to
the correlation, the wide apertures cause an intrinsic dispersion such
that particles of the same momentum reach the detector spread over a
tens of centimeters horizontally.  The dispersion also ensures that particles of the same
momentum are being measured by a range of physical scintillator strips in each plane.

The four wire chambers were originally built for the HyperCP 
experiment~\cite{Burnstein:2004uk} in the late 1990s.  The upstream two have an
aperture of 457 x 254 mm with a wire pitch of 1.016 mm in X,U,X',V
configuration with U and V rotated by $\pm$26.57 degrees.  The downstream two
are larger, with an aperture of 559 x 305 mm and wire pitch 1.270 mm.
The planes and original electronics were refurbished for our use.

The TOF units are used to measure the time the particle travels from
just in front of the first wire chamber to just behind the last wire
chamber.  The path length is 6.075 m with RMS variations of 0.014 m
from center to inside and outside tracks through the bend magnets.
The front TOF unit is a single piece of inch-thick
scintillator.  The back unit is three longer pieces of inch-thick
scintillator, covering an area larger than the wire chamber aperture.
A resolution of 200~ps is obtained using fast photo-multiplier tubes (PMT) reading
out two sides of each scintillator and a 25~ps least-count time to
digital converter.   The photo-electron yield,
scintillator size, and length of signal cables contribute to this resolution.

With the 1.07 m wide detector, large magnet and wire chamber apertures, and our
chosen beam tune, the beam delivers a broad distribution of
protons and pions from 0.35 to 3.0~GeV/c in momentum.  
The usable momentum range for these analyses is 0.35 to 2.0~GeV/c which
provides proton, $\pi^+$, and $\pi^-$ samples, each total roughly ten
thousand particles.
The electron content of the beam is small and limited to
momenta below 0.5~GeV/c, but has enough events that statistical
uncertainties are smaller than systematic uncertainties.  In addition,
there is a 5\% kaon component, plus smaller components of
deuterons and alpha particles which are not included in the
results presented here.

\begin{figure}[ht]
\begin{center}
\includegraphics[width=9cm]{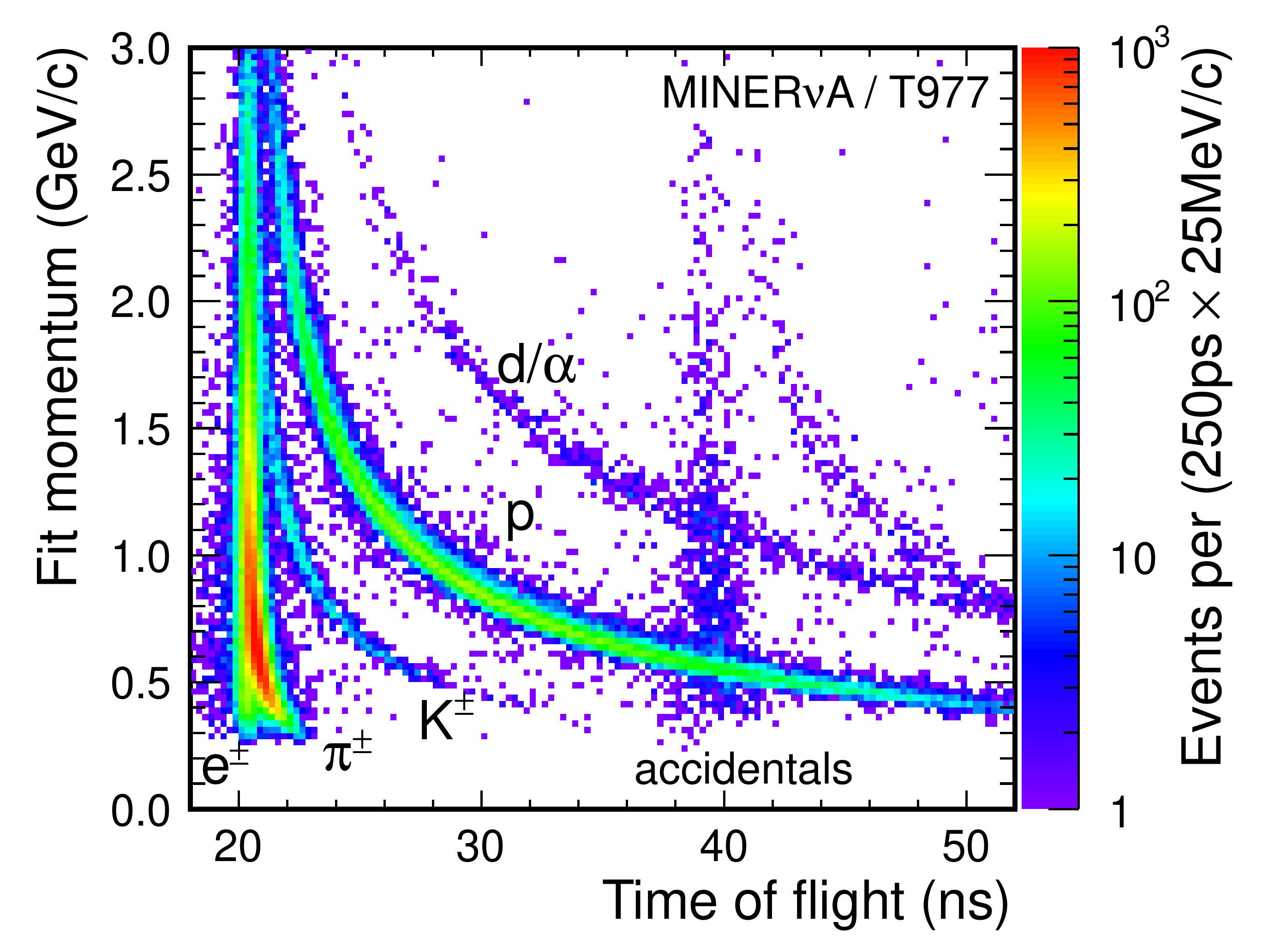}
\caption{ The measured momentum and time-of-flight used
    to separate different particle species and backgrounds. The origin
  of the backgrounds from the accelerator timing structure are described in the text.}
\label{fig:beamline}
\end{center}
\end{figure}

The pion, kaon, proton, and deuteron/alpha components are well
separated, shown in Fig.~\ref{fig:beamline} after quality cuts to
ensure only well-reconstructed particles.
Low momentum electrons are barely discernable near
20~ns in this figure.  There is also an accidental background near 39~ns when another
particle coincidentally passes early through the upstream TOF.  These
protons and pions happen because
the Fermilab Main Injector Accelerator supplying the beam has a 53 MHz
time structure.  Another pion striking the copper target earlier
than the triggered proton, kaon, or pion particle
can produce a particle that passes through the upstream TOF, creating
a timing artifact at integer multiples of 19~ns.

The separation shown in Fig.~\ref{fig:beamline} allows species to be
selected based on momentum, TOF, or the combination of the two plus
the measured path length that
gives an estimator for the mass of the particle.  Protons (and kaons)
are selected by requiring the estimated mass be within $\pm$20\% of the true mass
of the particle.  The selection is wider for pions and based on TOF because the TOF
resolution is the limiting factor and we do not cut events
if their TOF measurement fluctuates to superluminal.  The lower
bound is 
the expected TOF of a pion less 0.5~ns and rejects electrons; the upper end is based on
the TOF for the pion mass plus 20\% plus an extra 0.5 ns.  
An additional cut from 38~to~41~ns rejects the accidental pion background in the
proton sample.  The purities of the pion and proton selection are better
than 99\%, as estimated by extrapolating the tails of the wrong-species
distribution under the selected events, plus additional ab-initio
simulation of the electron content of the beamline design.
The electron selection is more complex and includes pion rejection criteria,
as described in Sec.~\ref{sec:electron}.

The momentum estimate is accurate to 1\% at low energy and 2\% at high
energy.  It uses a detailed map of the magnetic field
calculated using finite element
analysis software from the specifications for the two dipole magnet coils and
steel and the position survey of their placement relative to each other.  The
central value of the magnetic field from the calculation is adjusted
down 0.58\% to match the
actual field of the magnet from in-situ measurements. 
Measurements of the field were taken by
stepping a 3D Hall probe through the magnet along vertical,
longitudinal, and transverse lines with both magnets installed in
their final positions.  The field measurements are well described
by the calculated field.  The description of the principle component of the field along
an axis through the magnets, especially the longitude extent of the
field, is the most important feature constrained by the measurements
and contributes a 0.5\% uncertainty in
the momentum.  The other uncertainty comes from the
accuracy of the position survey of the four wire chambers. 

The momentum resolution is also evaluated particle-by-particle.  It 
is 2.5\% for pions and ranges monotonically from 5\% to 3\% for low to high momentum
protons.  It
is driven by multiple scattering and nonuniform magnetic field
effects at low momenta and by the wire pitch
and beamline length at high momenta.  
The iterative momentum fit steps along a candidate trajectory through the nonuniform calculated
field to estimate the field integral.  Then a Kalman filter technique
\cite{Fruhwirth:1987fm} is
used to obtain the momentum and its uncertainty for each particle.
The resolution of the momentum estimate is modeled accurately
enough and is not a limiting factor for these analyses.

\section{MINERvA test beam detector and calibration}

The detector exposed to this FTBF beam (hereafter called the test beam detector)
is a miniature version of the MINERvA detector installed in the NuMI
neutrino beam \cite{Aliaga:2013uqz} (hereafter called the MINERvA detector).  It is made of 40
square planes of 63 nested, triangle-shaped scintillator strips each with
length 107 cm and thickness 1.7 cm.  
This contrasts with the MINERvA detector which has a
hexagonal cross section and is made of 124 planes of 127
strips in the central tracker region followed by another 20 planes
each of electromagnetic calorimeter (ECAL) and hadron calorimeter
(HCAL) which have lead and iron interleaved respectively.  
Both detectors share the same three-view UXVX sequence of
planes with U and V rotated $\pm$60 degrees relative to the X plane
that defines (for the test beam detector) the vertical coordinate
system.    Three views allow for reconstruction of
multiple tracks for the MINERvA detector and very good reconstruction of single
tracks in the test beam detector.

Unlike the MINERvA detector, the test beam detector's removable
absorber planes allow us to take exposures in two configurations.  One has 20
planes with 1.99~mm thick lead absorber (ECAL) followed by 20 planes
with 
26.0~mm thick iron absorber (HCAL).  The absorber is interleaved
by placing one absorber upstream of each scintillator plane.
The other has 20 planes with no absorber (tracker) followed by 20 planes of ECAL.
For compactness, this document will refer to these configurations as
EH and TE, respectively.

\begin{figure}[ht]
\begin{center}
\begin{minipage}[t]{6.8cm}
\mbox{}
\includegraphics[width=6.8cm]{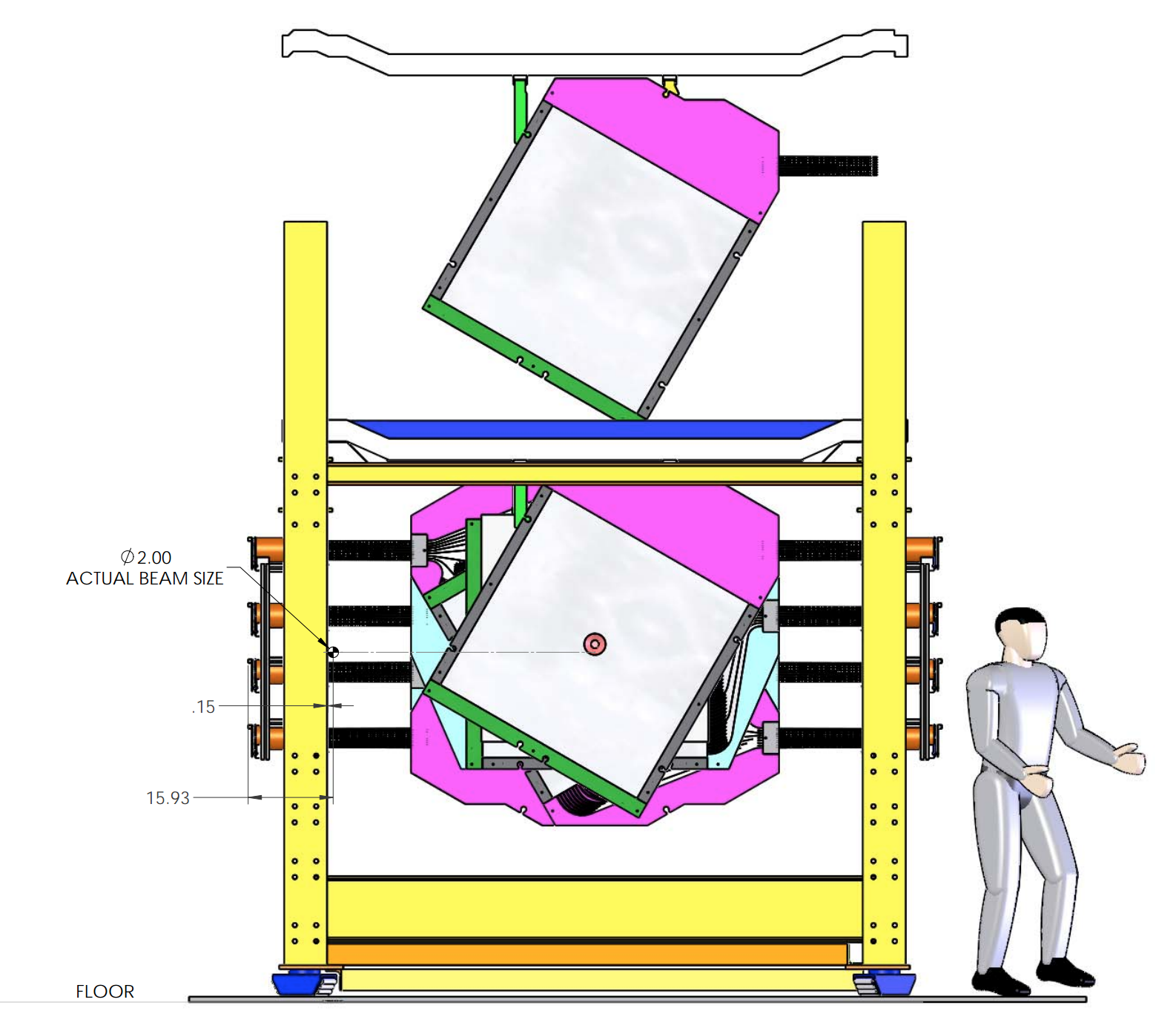}
\vfill
\end{minipage}
\begin{minipage}[b]{5.1cm}
\mbox{}
\includegraphics[width=5.1cm]{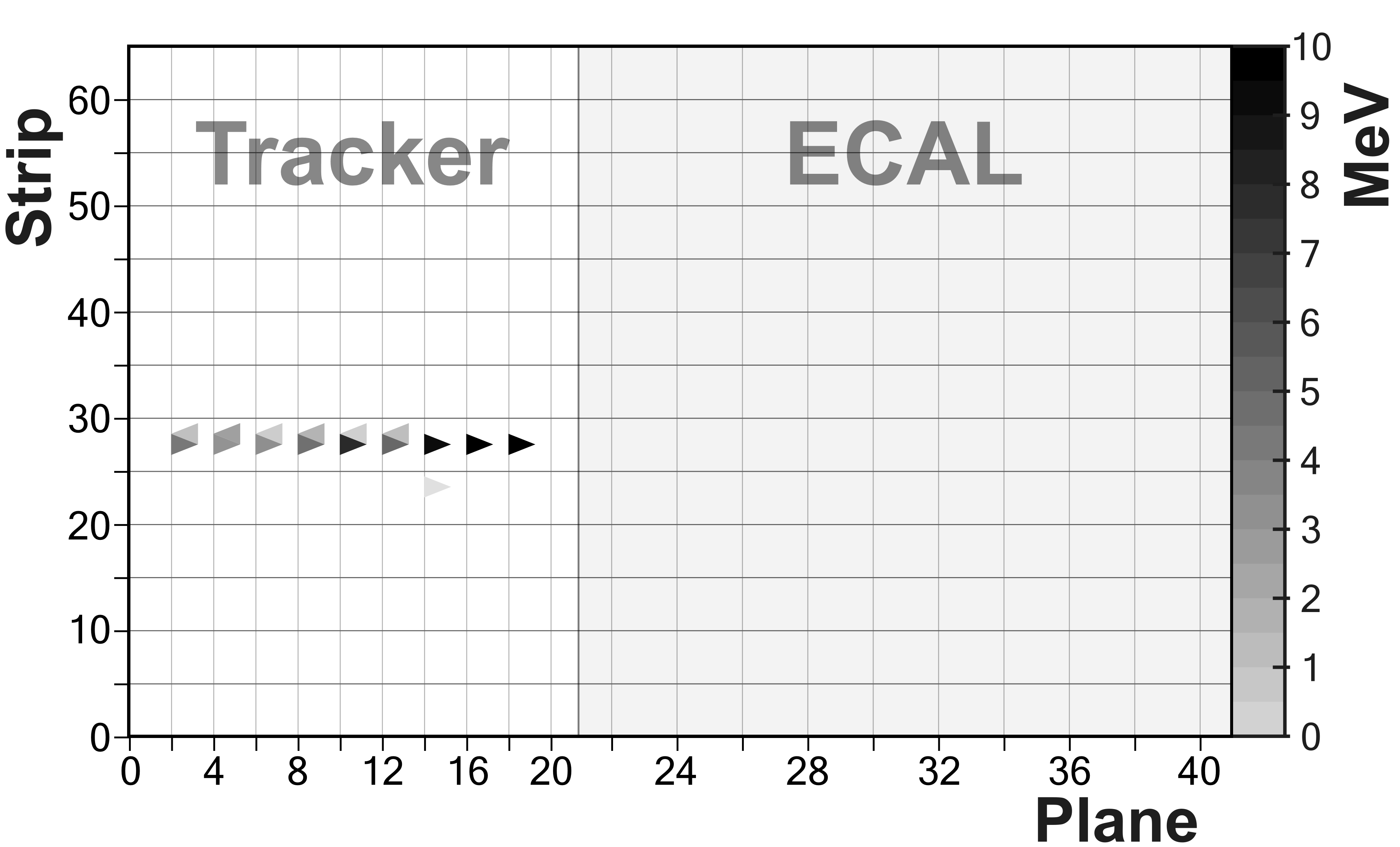}
\includegraphics[width=5.1cm]{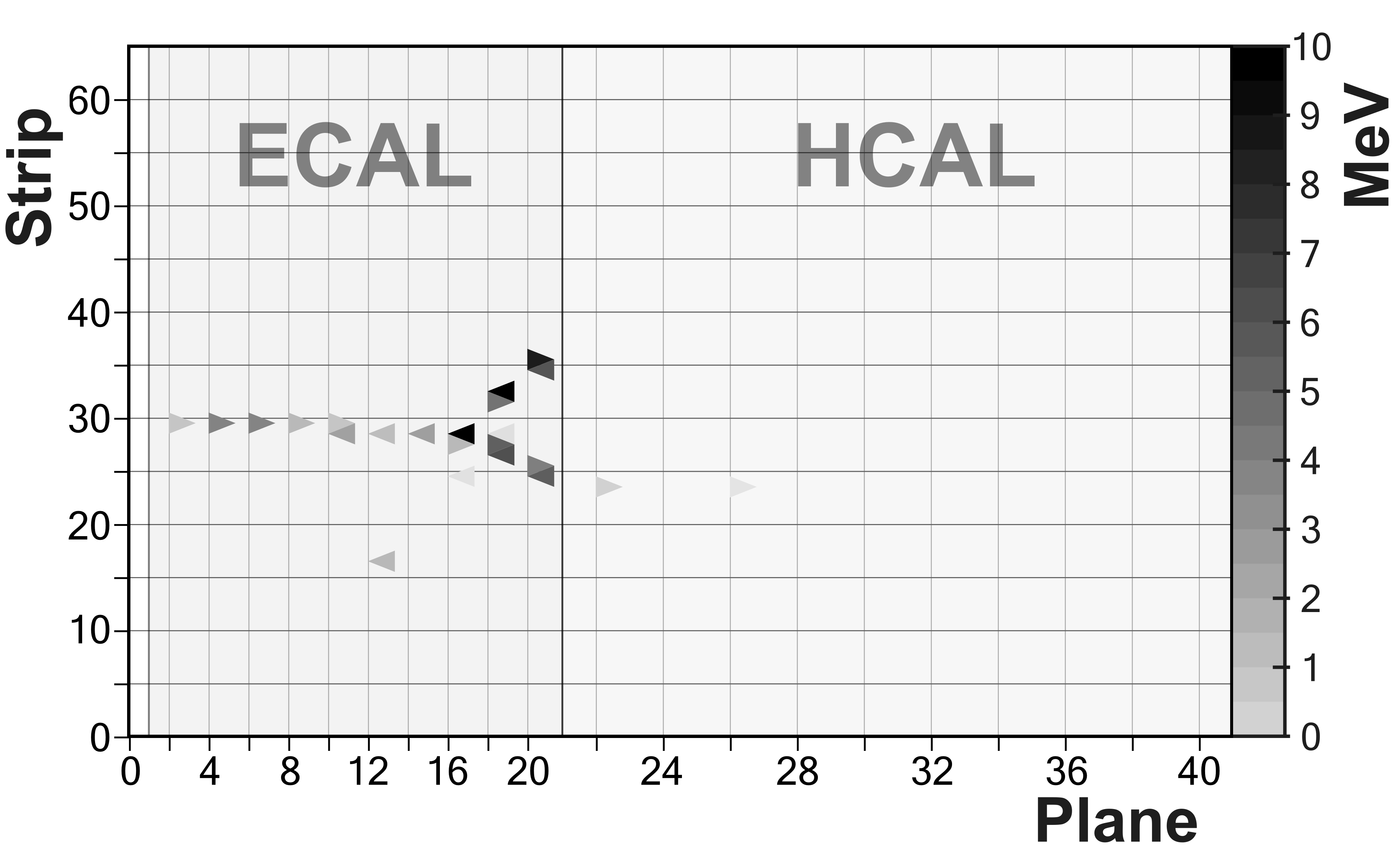}
\vfill
\end{minipage}
\caption{(left) A drawing of the detector viewed from the
  front.  The third U plane is being lowered onto the stand behind eight
  installed planes.  The drawing also illustrates the every other side
  readout in sets of four.  If this was the ECAL, there would be a
  plane of lead absorber between each plane.  A proton (right top)
  stopping at the end of its range in plane 18 of the TE configuration
  and a pion (right bottom) interacting near plane 16 of the EH
  configuration.  The aspect ratios for the right plots are not to
  scale and only activity in the X vew planes are shown. }
\label{fig:detectorpicture}
\end{center}
\end{figure}

As illustrated in the left panel of Fig.~\ref{fig:detectorpicture}, 
the first nine planes of the TE are shown with
no absorber.  Starting before the 20th plane, another hanger
with a sheet of lead would be lowered before each succeeding
scintillator plane.  For the EH configuration, a hanger holding a lead
sheet is installed before the first U plane and for all the first
20 planes.  Then a hanger with an iron plate is installed in front of
each of the remaining 20 planes.  The right panels are modified
from the web-based event displays \cite{Tagg:2011wk} for two events
from data. 
They show side view of
the X planes for a proton in
the TE and a pion in the EH detector configurations.
The design replicates the main downstream region of the MINERvA detector, which
has 124 planes of tracker followed by 20 planes of ECAL and 20 planes
of HCAL.  

The readout chain from scintillator to wavelength-shifting (WLS)
fiber to photomultiplier tube (PMT)~\cite{Aliaga:2013uqz} to digitization~\cite{Perdue:2012hg} is almost
identical between the test beam and the MINERvA detectors.  The exception
is that the test beam
detector has no clear fiber optical cables; the WLS fiber connects directly to the
PMT a half-meter out of the plane.  The effect of smaller scintillator planes and no clear fiber is that
the test beam detector has about 50\% higher light yield for a given
energy deposit, and
correspondingly better resolution for some kinds of measurements
compared to the MINERvA detector.
The Hamamatsu H8804MOD-2 multi-anode PMTs are the same.  The front
end electronics and DAQ~\cite{Perdue:2012hg} save the same 16~$\mu$s
of data in response to the trigger, and are only modified to operate
in response to a trigger formed by beamline instrumentation or cosmic ray
trigger scintillator rather than the predictable arrival of a trigger
from the NuMI beamline.

Unlike the MINERvA detector, in which the PMT assemblies for every
plane are on the same
side, on the test beam detector, the assemblies are alternated in groups of four
planes, one UXVX set rotated 180 degrees.   Mechanically this allows the planes to be
placed closer together than the frames and PMT assemblies would
otherwise permit.  The result is an air-filled space only slightly larger
than in the MINERvA detector.  Because the beam bend-magnets steer different
momentum particles to different portions of the detector (and at
different angles), there is a correlation between the geometry and the
position-dependent optical attenuation of the readout.  
Alternating the readout mitigates a few-percent momentum dependent
uncertainty, making this effect negligible. 

The detector energy scale is calibrated using the same strategy
described in \cite{Aliaga:2013uqz} for the MINERvA
detector.
An initial estimate for photoelectron yield is obtained for each strip
using pedestal subtraction and a gain measurement based on the light
injection system.   The intrinsic
differences in response between strips are analyzed using
through-going muons to produce a correction factor
to make the average response uniform from strip to strip.  As a side effect,
these muons give geometric plane position corrections.

The absolute energy scale is determined using a muon equivalent unit
technique.  The peak number of photoelectrons at the PMT is tuned to be the same in the data and
simulation, and the simulated geometry and Geant4 energy loss are
used  to set the absolute energy scale.  The calibration uses the peak of the
$\Delta$E/plane response for muons, and depends little on muon $\delta$-ray and
bremsstrahlung production in the tail of that distribution.
One difference between the test beam detector and the MINERvA detector calibrations
is the former uses broad spectrum cosmic ray muons, and a simulated
spectrum with the same angular distribution, rather than
momentum-analyzed muons from the NuMI beam.
As with the MINERvA detector, these calibrations do not include energy that
appears off the muon track due to crosstalk, a feature treated separately in the analyses
described in this paper.  

Temperature dependence is more important for the test
beam detector than it
is in the NuMI hall.  Hadron beam data were taken in June and July~2010, usually
during the hours from 04:00 to 18:00.  In addition to day-night
thermal changes, the heat load from operating the magnets and wire
chamber electronics warmed the experimental hall during the day, which then cooled
at night.  Thus, the overnight cosmic muon sample spans the same 23~C
to 34~C range of temperature as the daytime hadron sample.  
The detector response is corrected for the measured
temperature dependence on a plane-by-plane basis
and a residual uncertainty is included with the systematic
uncertainties.  Each plane is connected to a single PMT, but the
temperature dependence is the sum of effects due to the scintillator, WLS
fiber, and PMT.  Averaged over all planes, the effect on the energy scale
is -0.42~$\pm$~0.04~\%/C for the EH configuration and -0.37~$\pm$~0.03~\%/C for
the TE configuration.

\section{Data sample and simulation}
\label{sec:data}

\begin{figure}[ht]
\begin{center}
\includegraphics[width=6.0cm]{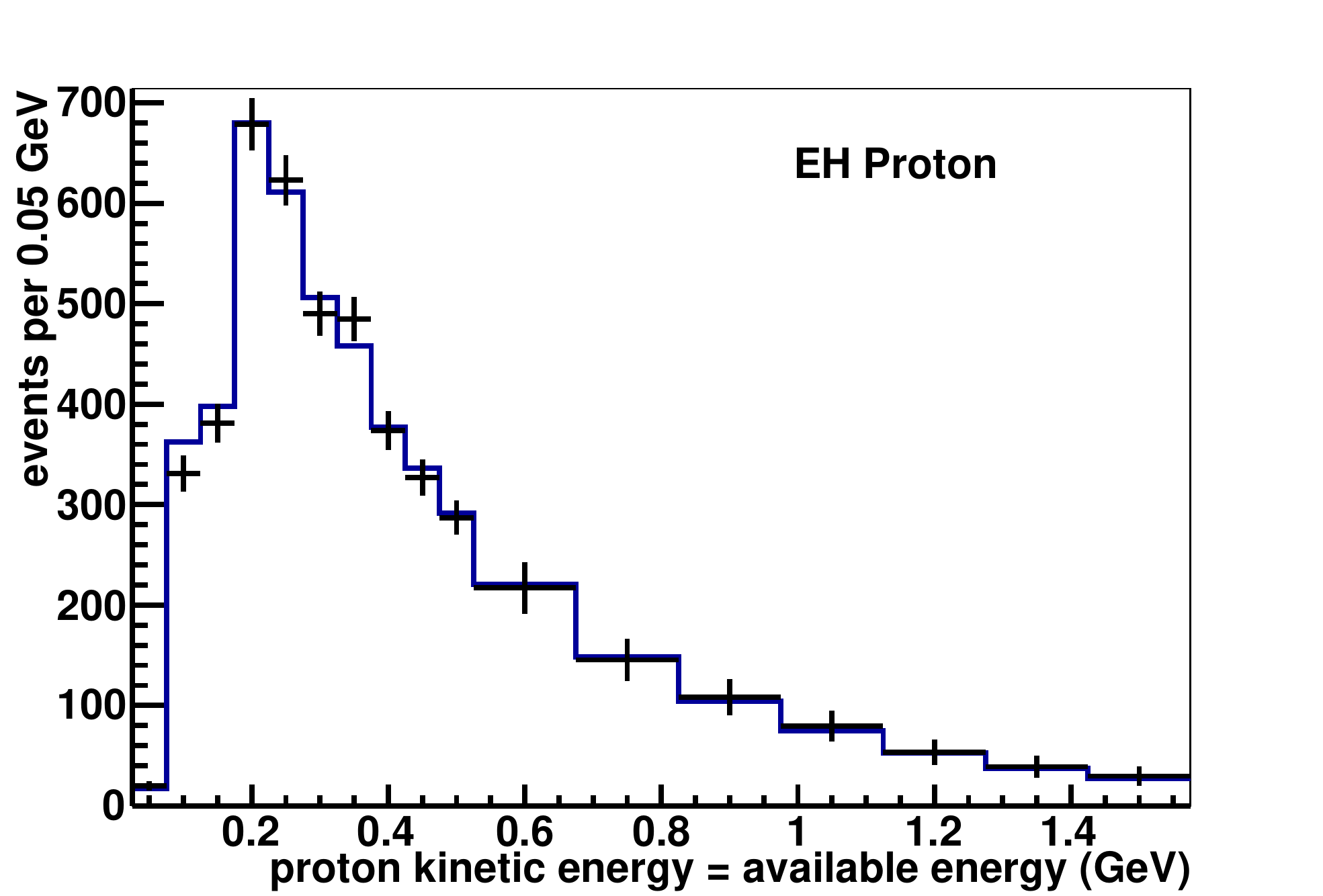}
\includegraphics[width=6.0cm]{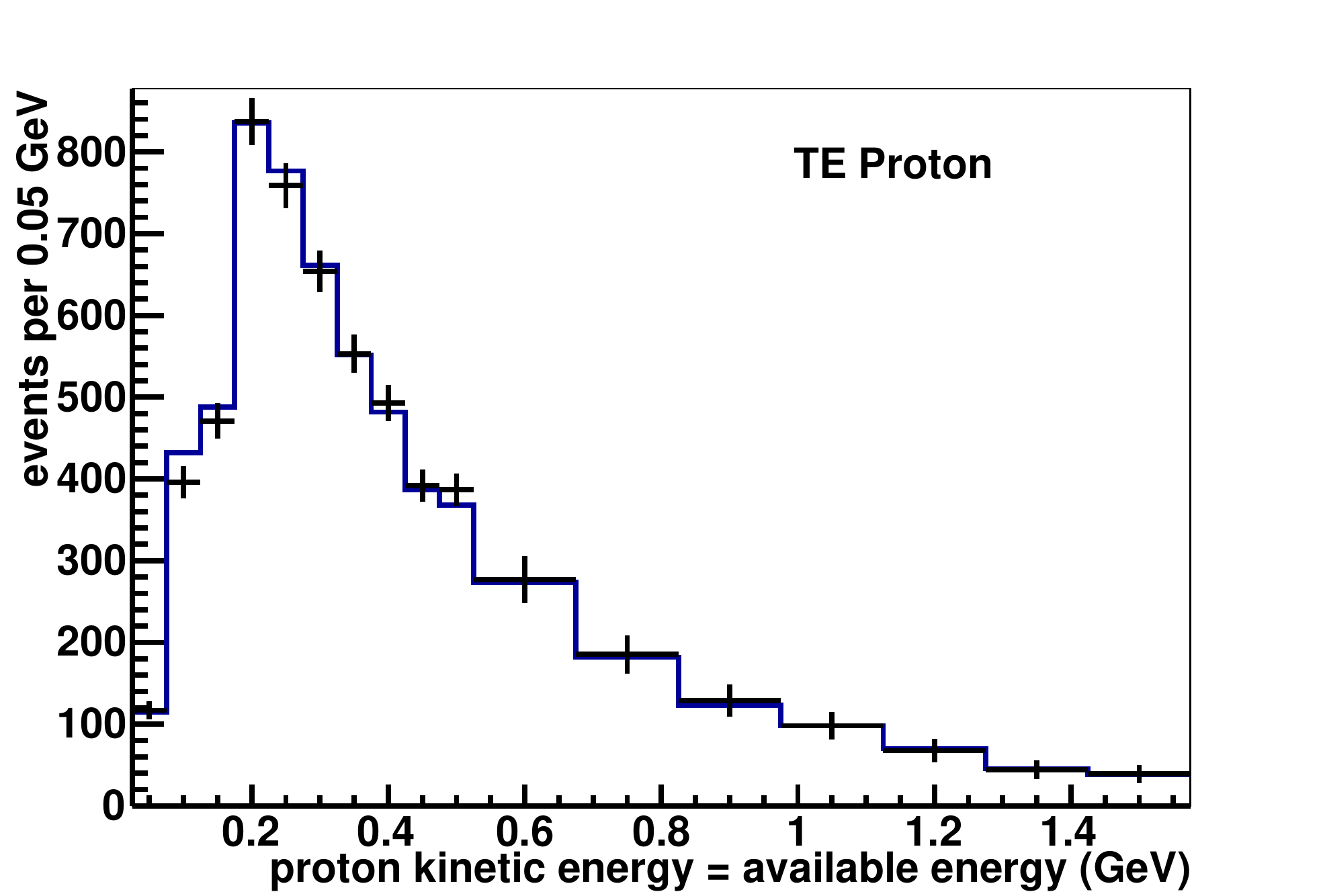}
\includegraphics[width=6.0cm]{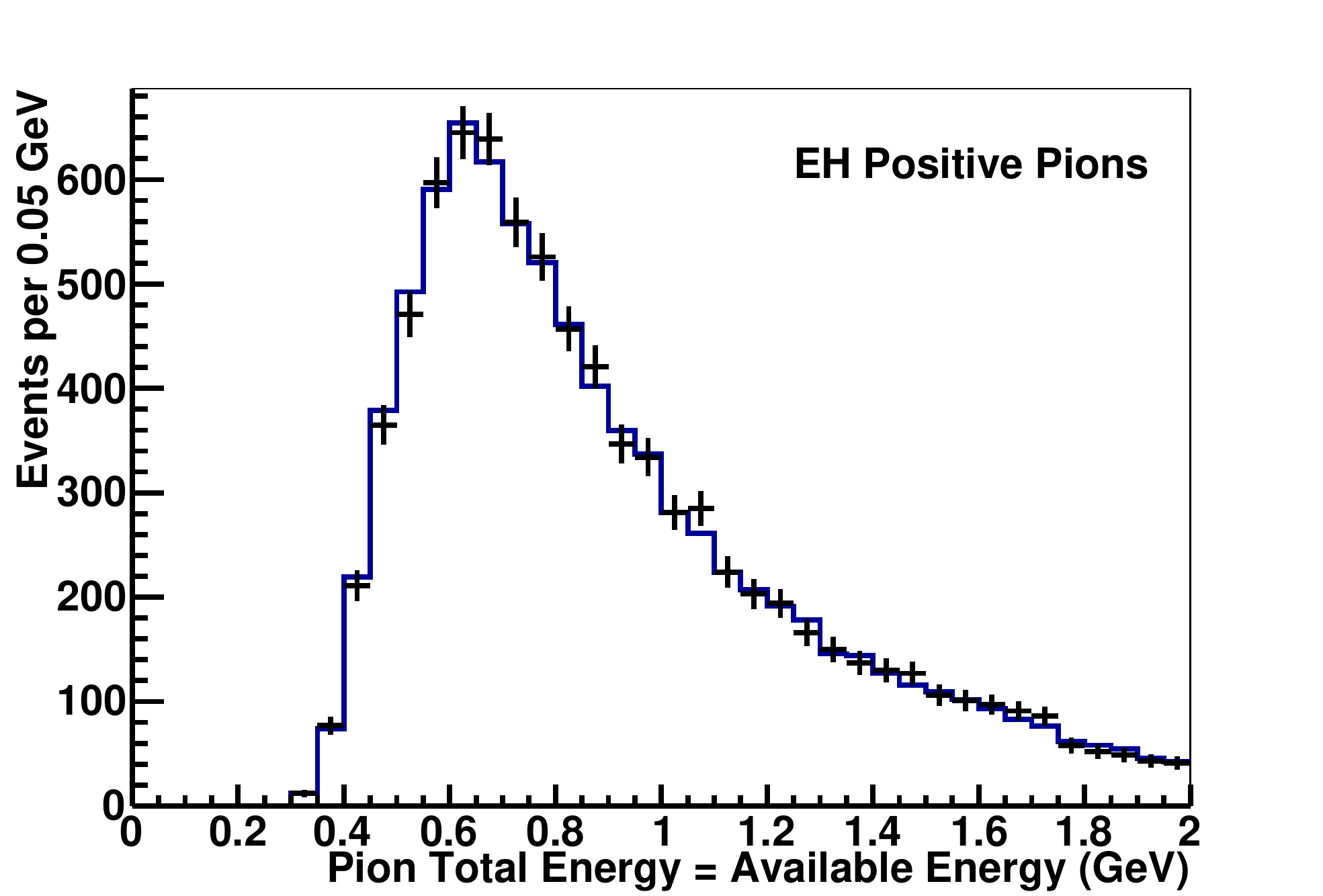}
\includegraphics[width=6.0cm]{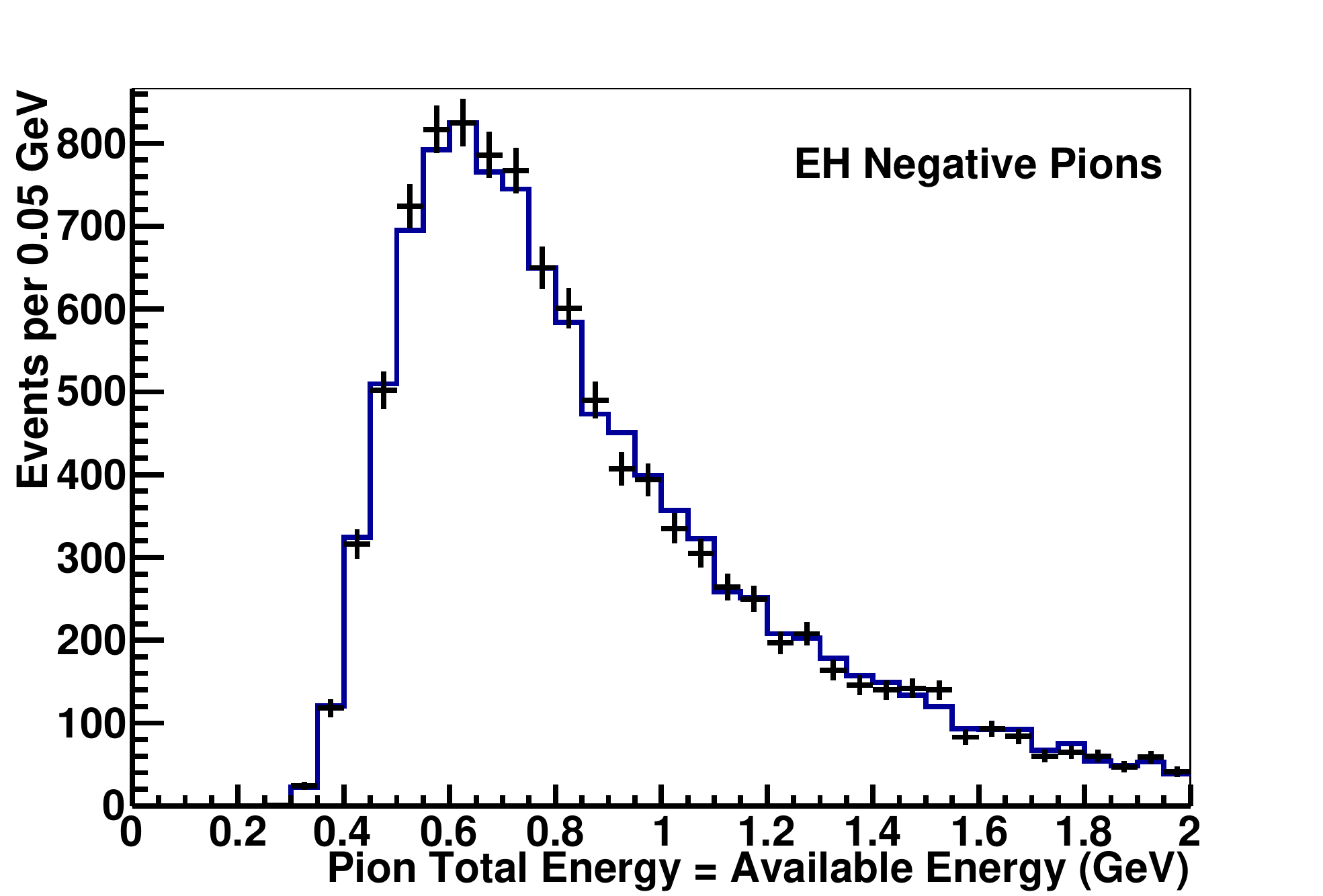}
\caption{ Measured spectra for EH proton (top left), TE proton
  (top right), EH $\pi^+$ (bottom left), EH $\pi^-$ (bottom right) 
 samples, after selection.  The histogram is taken from a Monte Carlo
simulation that was seeded with measured particle momenta and
trajectories from the data,
so by construction the spectra are the same.
\label{fig:spectra}}
\end{center}
\end{figure}

The incident particle momentum spectra for the selected data samples 
and the matching simulated spectra
are shown in Fig.~\ref{fig:spectra}.
There are plots of the energy spectra for protons in the TE configuration
and p, $\pi^{+}$, and $\pi^{-}$ in the EH
configuration.  At these
momenta, pions leave the back of the TE detector and are not used for
a calorimetric analysis.
The data samples are selected using the momentum and time-of-flight
measurements shown in Fig.~\ref{fig:beamline} and discussed in
Sec.~\ref{hadronbeam}.  The proton distribution is not smooth at
0.15~GeV because of the extra TOF-based pion-background rejection.

In this analysis, the data are compared
to a detailed, high-statistics Monte Carlo
simulation (MC).  The different species' spectra for the simulation are generated from
the {\em data}
particles'  position and momentum measured at the third wire chamber, with
momentum and angle smeared according to the
estimated resolution on a particle-by-particle basis.  The simulation then
propagates particles through the material of the third and fourth wire chamber, the downstream TOF, the cosmic muon
trigger scintillator, the air, and finally into the test beam
detector.    Using the estimated
energy and position resolution for each particle, we apply a Gaussian random 
smearing and use the same initial particle position and momentum
multiple times.  The resulting MC samples are typically 20 to 40
times larger than the data, depending on the analysis.  In our limit
of excellent resolution, this method is adequate to replace a full
unfolding of the resolution.

The MC does not include any beamline-induced background effects, neither from
particles that are exactly in-time (from the same parent 16~GeV pion
hitting the target) nor from secondaries from another pion in a nearby 19~ns slot in
the Fermilab Main Injector 53~MHz accelerator structure.  
Because activity is saved from 16~$\mu$s
around each trigger, and because some incident particles
should spatially leave much of the detector quiet, the data contain a record of
the average beam-induced background around valid triggers. 

The particle selection technique was developed and validated using the
web-based MINERvA event display \cite{Tagg:2011wk},
in many cases with the help of undergraduate research assistants.
The main selection requires the particle to appear in the detector at
a location and time predicted by the measurement in the beamline.
Events with substantial unrelated activity, especially if it is
track-like, are rejected.  There is about a 10\% background mainly
from muons, but also lower energy particles entering the detector with
the triggered event.  
Events with additional reconstructed activity within 250~ns before and 500~ns
after are also removed.
These selections reduce beam-induced backgrounds and also
eliminate triggered particles that scattered substantially in the
beamline before reaching the detector.   The selections to reduce
these unwanted events are applied to both the data and MC samples.
Finally, we estimate and make a
statistical subtraction of the remaining background and evaluate an
uncertainty specific to each analysis.

The basis of the simulation uses Geant4 version 9.4p2 \cite{Agostinelli:2002hh,Allison:2006ve} and
our best description of the detector geometry and material
\cite{Aliaga:2013uqz}.  
A summary of the material properties of each of the three detector
regions are given in Table~\ref{tab:material}.
The scintillator plane is made of 1.801~g/cm$^2$ of plastic scintillator,
WLS fiber, and a co-extruded TiO$_2$ reflective coating.  Added to
this is another 0.226~g/cm$^2$ of epoxy and Lexan. The
scintillator planes were made at the same facilities immediately
following the production of MINERvA planes, and the
modifications for assembling smaller planes make negligible
difference.  The uncertainty on the amount of the material in the
assembled scintillator planes is the same
1.5\% as for the MINERvA detector.  In the
ECAL portion of the detector there are planes of lead with 
thickness 2.30~g/cm$^2$ and in the
HCAL version there is 20.4~g/cm$^2$ material
that is 99\%~Fe and 1\%~Mn.  
The lead and iron absorbers are similar to those in the MINERvA detector, 
but we use the as-measured test beam detector quantities in the
simulation and to evaluate material assay uncertainties
(coincidentally) of 1.2\% for each kind of absorber.
The uniform, simulated air gap from one plane to the next is an approximation to the as-measured
air gap, and the absorbers are
approximately in the middle of this air gap, in front of the
associated scintillator plane.
The air gaps and also the approximate nuclear interaction and radiation lengths
are summarized in Table~\ref{tab:material}.
\begin{table}[ht]
\begin{center}
\begin{tabular}{l|ccccc}
   & material & percent & air gap & interaction & radiation \\
   &  g/cm$^2$ & uncertainty  &   mm   & lengths & lengths \\ \hline
Tracker & 2.027 &  1.5\% & 2.2& 0.5 & 0.9 \\
ECAL & 2.30 + 2.027 &  1.2\% + 1.5\% & 8.1 & 0.7 & 8 \\
HCAL & 20.4 + 2.027 & 1.2\% + 1.5\% & 3.5 & 3.6 & 30 \\
\end{tabular}
\caption{ Summary of the as-simulated material composition for each
  detector region.
The nuclear interaction length and the radiation length are for twenty
planes, the others are per-plane.
\label{tab:material}}
\end{center}
\end{table}

Almost all aspects of the detector response are simulated
using details constrained by calibration data and bench tests, including
Birks' law parameter measured from these data, described next in
Sec.~\ref{sec:birks}.  
The temperature correction is made to the data but not simulated.
Crosstalk arises because each scintillator strip's light is directed
onto a pixel in a 64 channel PMT, leading to optical and some electronic
crosstalk, which is simulated and tuned to data.  A few
features are not simulated, of which PMT after-pulsing and PMT
nonlinearity are the only significant ones, and only affect high pulse-height activity.

\begin{figure}[ht]
\begin{center}
\includegraphics[width=5.5cm]{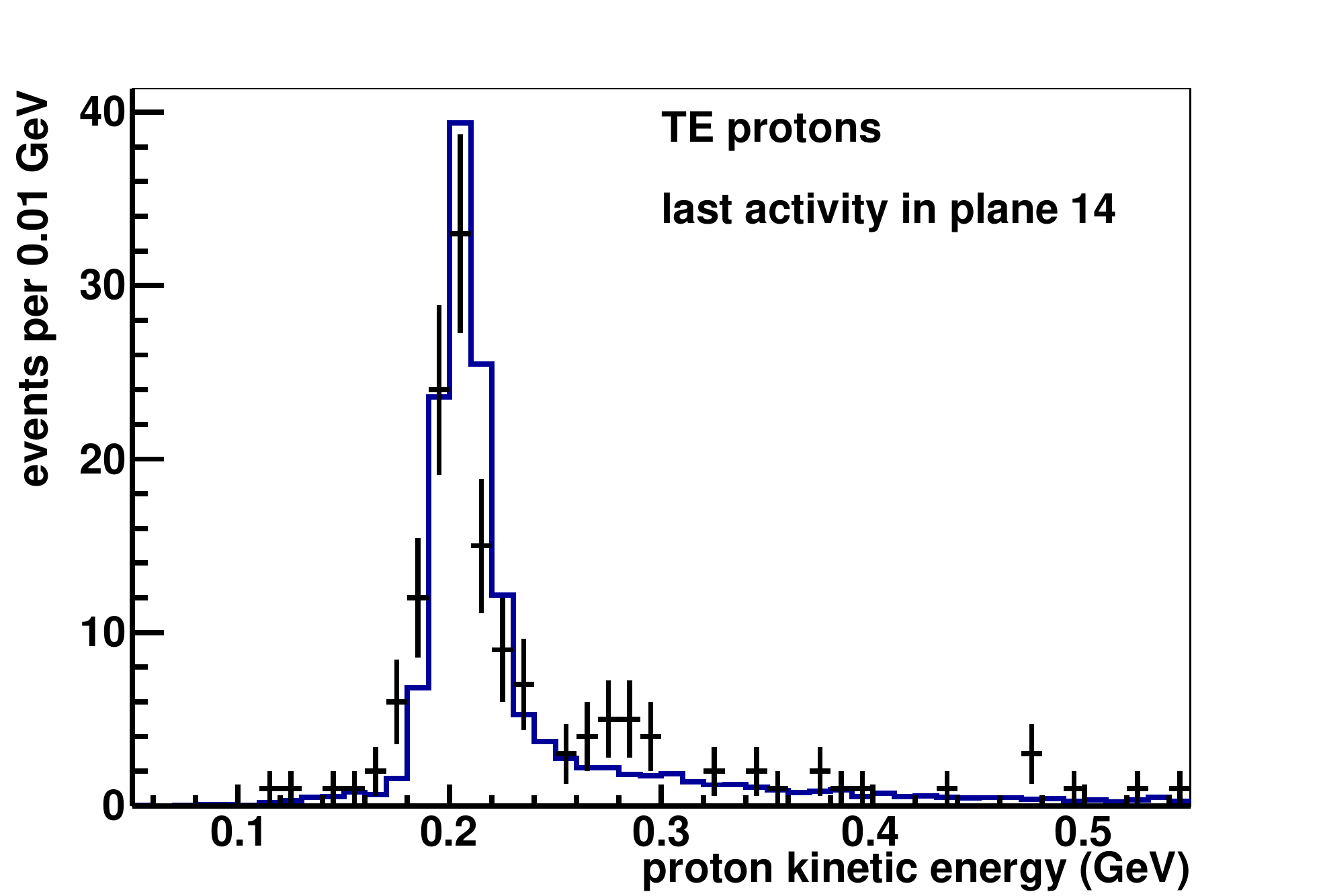}
\includegraphics[width=5.5cm]{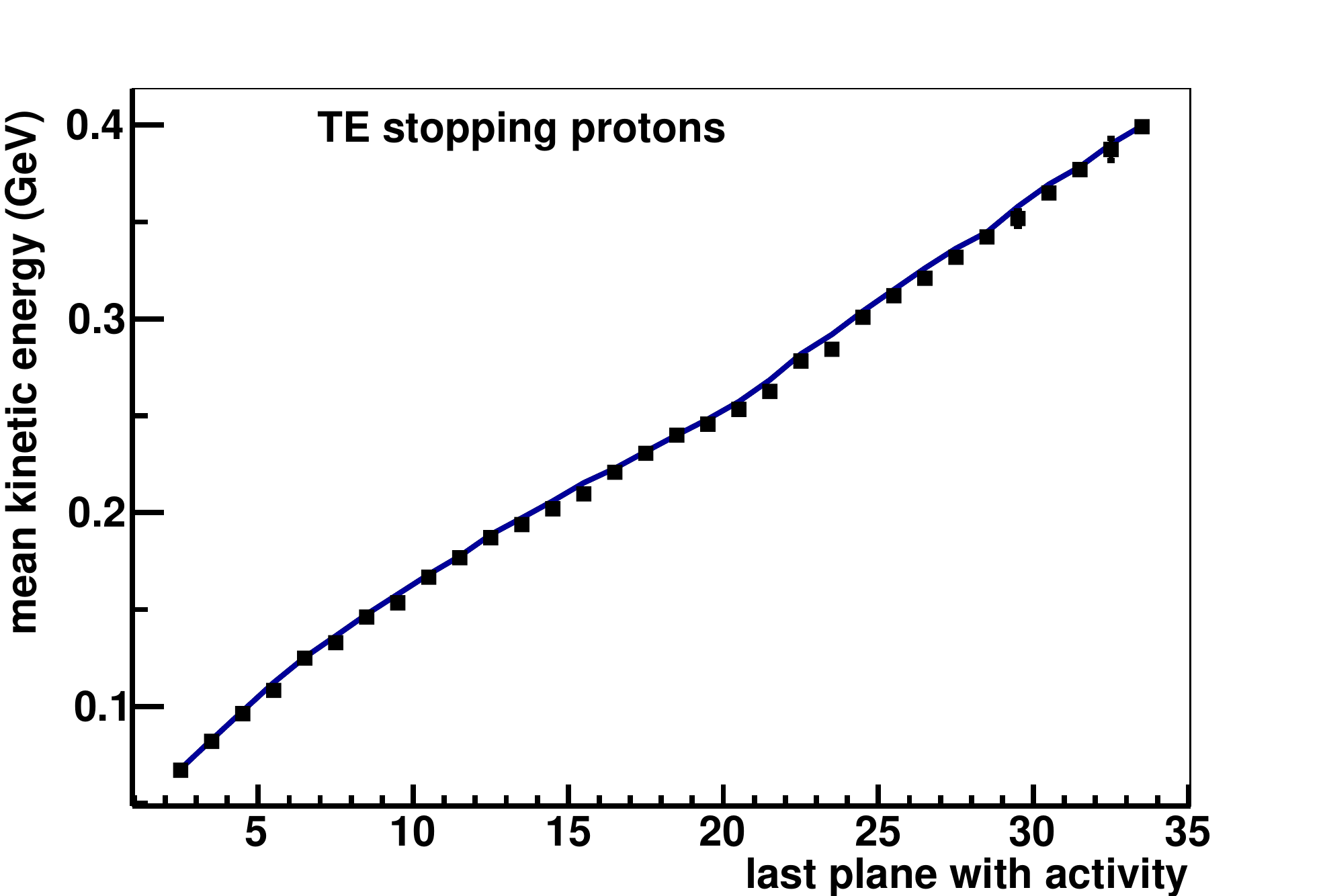}
\caption{ The left plot is the kinetic energy distribution for protons
  that stopped in plane 14 in the TE configuration.   Protons at the end
  of their range form a peak at 200 MeV.
  The right plot is the energy from the mean of a Gaussian fit to the peak for
  protons that appear to stop in each TE plane for data and
  MC.  The MC protons stop 1.1\% earlier than the data, a discrepancy which is smaller than the
  beamline momentum and material assay uncertainties.  Most error bars
  are less than 1\% and are too small to see.
\label{fig:energytomodule}}
\end{center}
\end{figure}

Proton range and proton single-particle energy resolution obtained
with the MC give good description of the data,
confirming that the beamline characteristics are well simulated.
To study this, the sample of protons in the TE configuration
is divided according to which was the furthest
downstream plane with activity.  The distribution of measured proton kinetic
energy for protons that make it to plane 14 (an X plane) is shown in the left plot of
Fig.~\ref{fig:energytomodule}.  The distribution has a peak near 200~MeV
corresponding to protons that were really at the end of their
range, and a high energy tail from protons that experienced an
interaction.
The protons that are actually at the end of their range form a
Gaussian-like peaked distribution which can be fit to obtain a mean energy
and a resolution.  This same sample is used later to select
stopping protons for the Birks' parameter measurement in
Sec.~\ref{sec:birks}, and potentially trackable protons for Sec.~\ref{sec:tracking}.
The procedure is done separately for both data and
simulated events; in neither case does selecting the subset of stopping protons
involve a prediction of the range.  

The proton range is 
well-modeled by the simulation.   The simulated protons stop 1.1\% earlier than
the data, which is a smaller discrepancy than the 1\% beamline momentum
plus 1.5\% material assay uncertainties.  A comparison of the 
Gaussian fit mean from the end-of-range peak 
is shown in the right plot of Fig.~\ref{fig:energytomodule} for data and MC.  Stopping
protons are such a high resolution sample, the widths of
those Gaussian fits (10 to 15 MeV, not shown) are primarily driven by the beamline and multiple
scattering resolutions, not effects of the test beam detector, and are also accurately described by the simulation.

\section{Birks' law parameter}
\label{sec:birks}

Birks' law describes the quenching
effect on scintillation photons produced by high, localized energy
deposits. 
After calibration of the beam and detector, we measure the Birks' law
parameter \cite{Birks:1951,Birks:1964zz} for the MINERvA polystyrene
scintillator \cite{Aliaga:2013uqz}. 
Birks' law quenching is an important effect at the end of proton tracks
and affects calorimetry measurements in the MINERvA and test beam
detectors.  A large sample of proton energy deposits at the end of
their range is obtained using the selection described at the end of
the previous Sec.~\ref{sec:data}.  We use the subset of events that appear to
stop in planes 9 to 19 of the TE configuration
and are in the peak of the distributions illustrated by Fig.~\ref{fig:energytomodule}.

\begin{figure}[ht]
\begin{center}
\begin{minipage}[t]{6.9cm}
\mbox{}
\includegraphics[width=6.9cm]{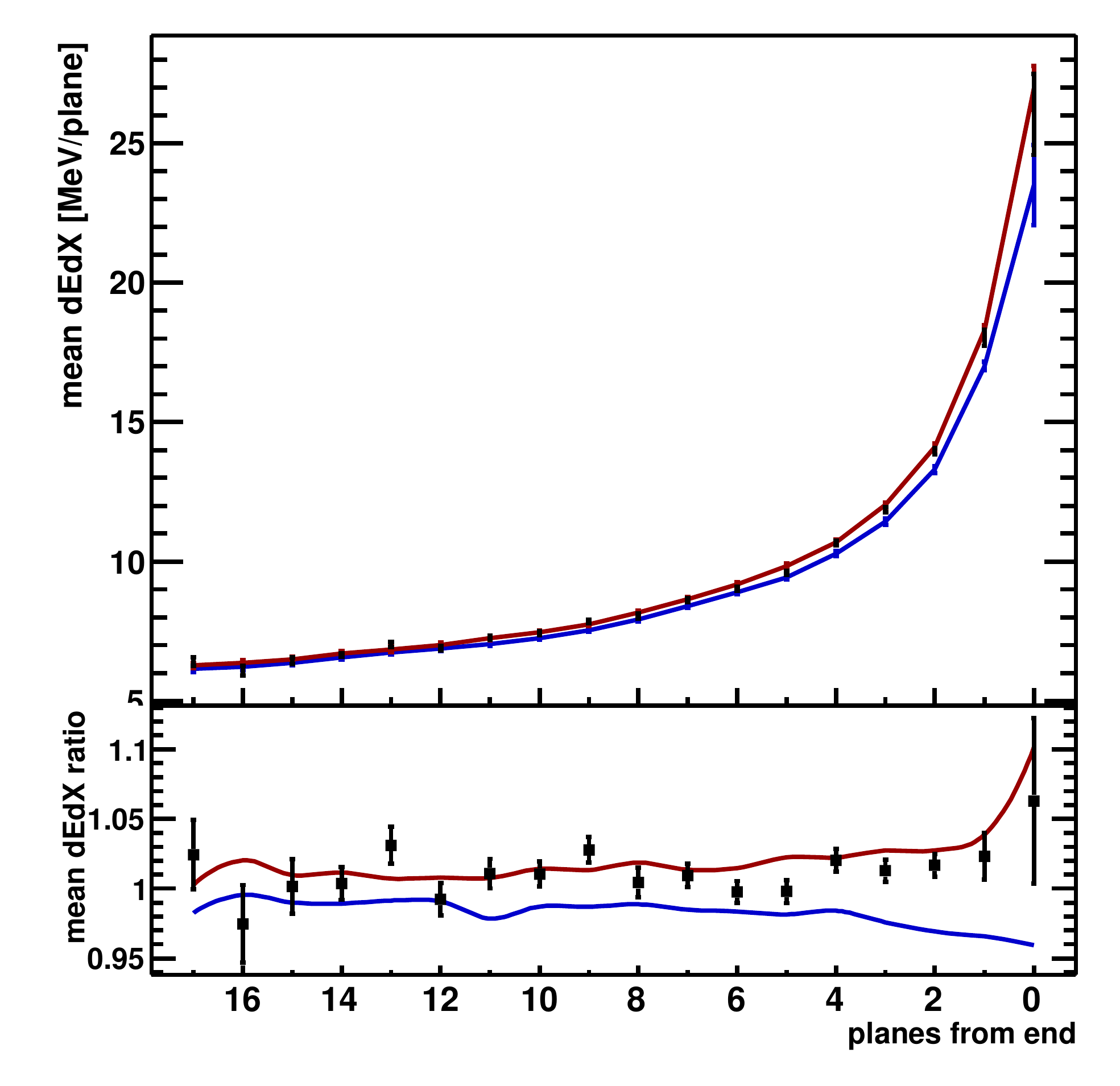}
\vfill
\end{minipage}
\begin{minipage}[b]{5.1cm}
\mbox{}
\includegraphics[width=5.1cm]{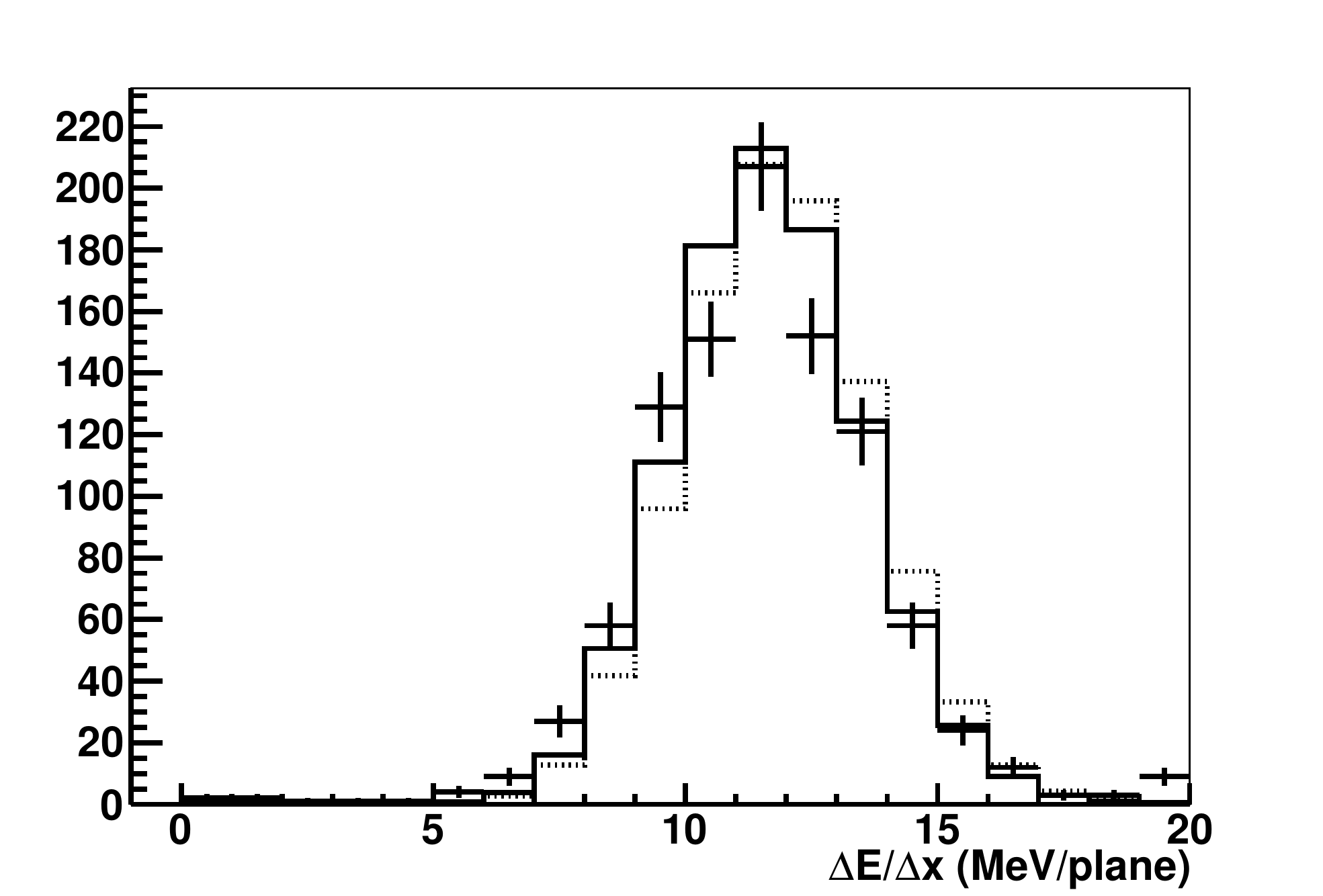}
\includegraphics[width=5.1cm]{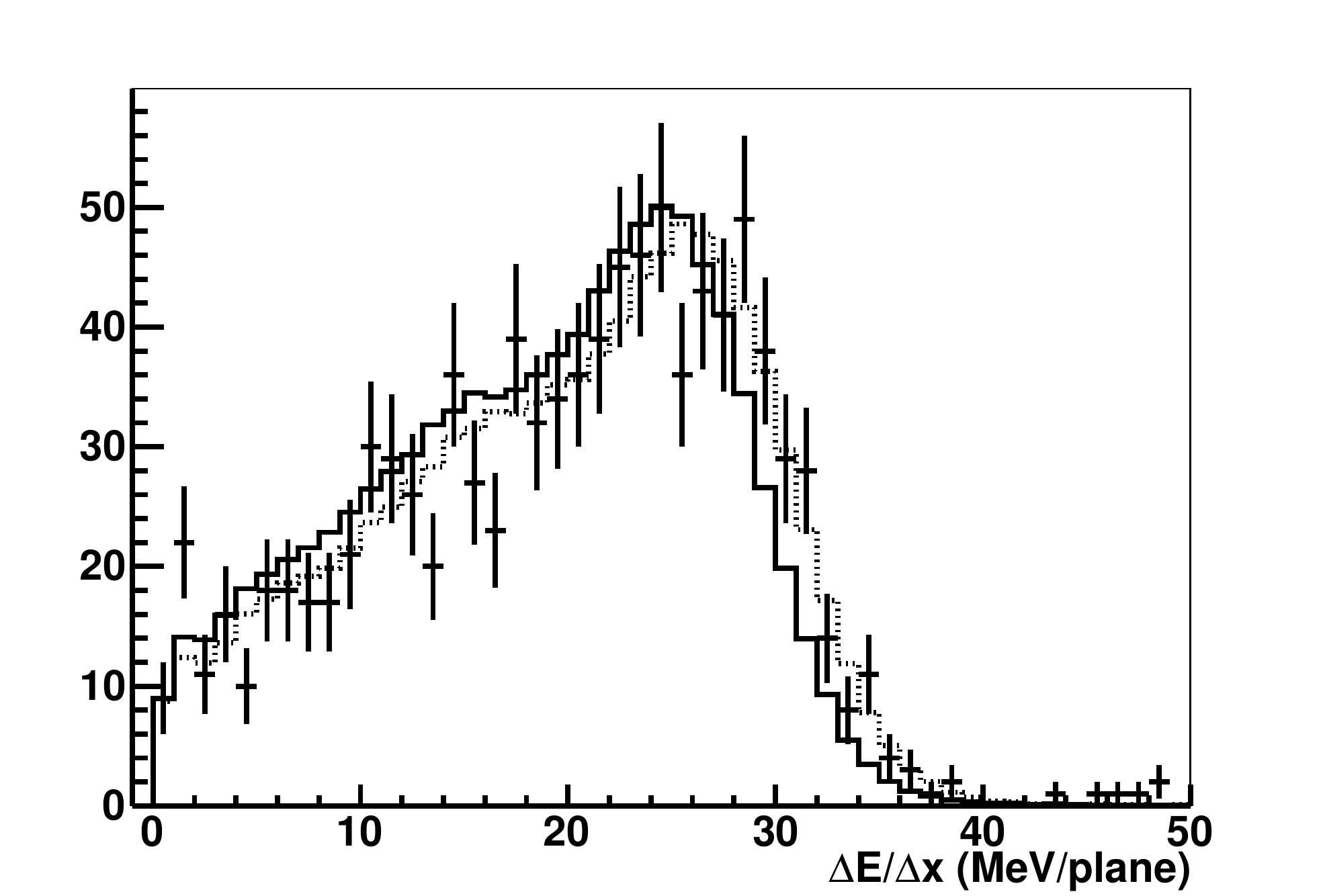}
\vfill
\end{minipage}
\caption{ The left plot shows the measured energy deposit per plane for data
    compared to the simulation with the before-fit Birks' parameter of
    0.133 mm/MeV and the original -30\% (top line) and +30\% (bottom line)
    uncertainty on this parameter's initial value.
    The figures on the right show the underlying binned energy per
    plane, {\em at best fit} for the non-Gaussian end plane zero
    (lower right), and the more Gaussian
    plane third from the end (upper right).  The MC distributions on the right show the smaller 17\%
    uncertainty bounds, one shifted to higher values on the horizontal
    axis, one shifted lower, such that the best fit parameter (not shown)
    would lie between them in every bin, see discussion.
\label{fig:birksprofile}}
\end{center}
\end{figure}

Birks'  empirical parameterization of the quenching factor to be
applied to photons/MeV is
$$\mathrm{Suppression\; factor}\; = \frac{1.0}{ 1.0 + \mathrm{Birks\; Parameter} \times (dE/dx)},$$
with one parameter, often abbreviated $k_B$ with units of mm/MeV.
This suppression is implemented in the MC and
applied to MC deposits based on the $\Delta$E and $\Delta$x as the
simulation steps the particle through the active scintillator material.  If the
parameter $k_B$ is too high, the MC will show a discrepancy of too
much suppression in the energy per
plane that increases toward the end of a proton's range, with the data
having the higher energy response.  The left plot in
Fig.~\ref{fig:birksprofile} shows such a trend using the default
value of 0.133 $\pm$ 0.040 mm/MeV used by MINERvA until the present
measurement.   The mean energy loss is better described by the top MC
line with lower parameter value and higher
response as a function of the distance from
the observed end of the proton's path into the detector.

The left plot of Fig.~\ref{fig:birksprofile}, and the extraction of a better value for Birks'
parameter, is formed using the binned distributions of the energy
deposited per plane.  Two
examples of the underlying data are shown on the right.
The top one is for the third plane from the
end, and one for the plane at the end (zero planes from the end).  
The plane at the end is the most sensitive to Birks' parameter, but
does not have a Gaussian shape, requiring a more complex fitting
technique than simply fitting the mean of the energy per plane.
These plots are used here to describe how the fit is constructed
from the binned data for each plane.
In the two distributions on the right, the two MC lines shown in each
plot are for the smaller uncertainty 0.0905
$\pm$ 0.015 mm/MeV at best fit.

Birks' parameter is extracted iteratively.  The original default value of the
parameter and its uncertainty are used to make three full MC samples to
extract a new parameter and smaller uncertainty.  As shown in
Fig.~\ref{fig:birksprofile}, the MC samples with $\pm1\sigma$ around
the default value of $k_B$ usually
bracket the data.  For a trial Birks' parameter, the predicted binned
distribution is formed by interpolating between these two samples, or
when necessary extrapolating slightly beyond these samples.  By scanning
through a full range of parameters, the one that returns the lowest
$\chi^2$ is used to seed the next iteration of the analysis.

Not all available bins or planes are used in the fit.  
Planes further from the plane in which the proton stopped
than the first fourteen are excluded from the
analysis.   They have low statistics because protons which only go nine
planes into the detector do not contribute.
To support systematic uncertainty studies, the analysis keeps only
bins in the central region of each distribution in the right plots in
Fig.~\ref{fig:birksprofile}.  This ensures all bins in the fit will remain populated when
systematically shifted samples are constructed.
In the example of plane three, bins from 6.0 to 15.0 MeV are included, 
while plane zero includes bins from 10.0 to 32.0 MeV.
In total, there are 123
bins across the fourteen planes-from-end included in the analysis.

The overall energy scale is an unconstrained parameter in the fit,
which simultaneously 
accounts for both the uncertainty in the energy scale and the correlation
between the calibrated energy scale and Birks' parameter.
Every iteration of the fit scans over this parameter by applying
a scale factor to each energy deposit of each MC event and reforming
each histogram.  
The scale factor causes individual entries in the plots on the right side of
Fig.~\ref{fig:birksprofile} to shift
higher or lower along the horizontal axis.
This is equivalent to moving the mean of each distribution by the same amount,
and allows the fit to seek a better $\chi^2$ minimum.

Also, an
amount of additional fluctuations of the simulated reconstructed energy
is allowed to vary from strip to strip.
This accounts for calibration
effects beyond those that are explicitly in the simulation or removed
from the data using calibrations.   At best fit, the analysis yields
the same additional 5.5\% smearing 
as found for the MINERvA
detector.  This parameter was not changed for every iteration of the
other parameters, only for values near
the best fit result in the later iterations.

In summary, the best fit value is obtained
using a parameter scan in this three-parameter space of Birks' parameter,
energy scale, and smearing of reconstructed energy deposits.  
The procedure is iterated with a new MC built from the new parameter
value and smaller $\pm$ shifted values.  After three iterations the procedure reliably
converges to the final result.

The best value for the Birks' parameter is 0.0905 $\pm$ 0.015 mm/MeV.
This value is near the
-1$\sigma$ limit of the original estimate used by MINERvA for
analyses through 2014, confirming that we used suitable Birks' effect uncertainties in
prior publications.  Future simulations using the  new value have half
the prior uncertainty.   The best fit describes the data well, yielding a $\chi^2$ of 124 for
120 degrees of freedom.  In addition to showing the method of the fit,
the two right figures in Fig~\ref{fig:birksprofile} show examples of the goodness-of-fit.

The measurement of the Birks' parameter is dominated by systematic
uncertainties.  One of the largest is from the correlation
with the energy scale, which is treated as an unconstrained parameter
in the fit.  The fit value is sensitive to variations of
which protons, which physical planes, and which bins are
included in the fit.
Uncertainties in the material assay are propagated to the result using
modified full MC samples.  Extra smearing of the scintillator and
PMT response to single energy deposits in the MC has a small effect.
Two special sources of uncertainty, the effect of Geant4 step size and
of PMT nonlinearity, are described below.  All these effects are
summarized in Table~\ref{tab:birkssystematics}.
\begin{table}[ht]
\begin{center}
\begin{tabular}{lc}
Source & uncertainty \\ \hline
uncertainty from fit & -7\% +5\% \\ 
proton selection & -11\% +3\% \\ 
Geant4 step size &  -0\% +9\%    \\  
PMT nonlinearity & -3\% +0\% \\
material assay & $\pm$5\% \\
physical planes & $\pm$5\% \\
MC energy smearing & $\pm$3\% \\
choice of bins & -3\% +0\% \\ \hline
Total & +16\%  -13\%
\end{tabular}
\caption{ Percent systematic uncertainties on the value for Birks'
  parameter from different sources.
\label{tab:birkssystematics}}
\end{center}
\end{table}

Birks' parameter is an effective parameter because it is obtained by
matching MC to data.  In addition to
describing the quenching of scintillation light, it is accounting
for the accuracy of the Geant4 energy loss simulation and our choice
to use the default (adaptive) Geant4 step size.
Allowing Geant4 to take more coarse steps,
up to one scintillator bar per step, yields an increase in the
simulated response of about 4\% in the last plane and a slightly better
$\chi^2=118$.  The typical simulated $\Delta$x has increased,
so $\Delta$E/$\Delta$x has decreased, so there is less Birks'
suppression applied.  This large variation
would cause a bias in the fit Birks' parameter of 9\%, about
half the total uncertainty.  However, this particular measurement is
specifically matched to the settings for
Geant4 that are used by the MINERvA
simulation as of late 2014.  This uncertainty should be included when
comparing to other measurements but is not an uncertainty on the
resulting simulation used for MINERvA neutrino analysis.

The PMTs have a nonlinear response due to saturation effects that
increase with dynode current and therefore total charge, an effect
separate from the ADC to charge calibration.  Nonlinearity has 
systematic effects on calorimetry and on Birks'
parameter because it is unsimulated and uncorrected.   
This nonlinearity sets in for high instantaneous current
at the anode, and so is a function of charge measured by the front
end board's digitization circuitry.  The result is a suppression: the
response in the data will be
be systematically lower than the equivalent MC events.  
MINERvA does not have a measurement under circumstances that are the
same as the 
light propagating in our scintillator bars and WLS fiber.
Instead, we have a reference nonlinearity curve informed by bench
tests in which the suppression is
parameterized by a quadratic function of the integrated current.

Because some light reflects from the mirrored far end of the WLS fiber
and reaches the PMT at a later time, our ab-initio upper bound on the
amount of suppression is taken to be half the reference amount;  this
is the baseline uncertainty.
The $\Delta$E/plane profile in Fig.~\ref{fig:birksprofile} is distorted by
nonlinearity in different ways from either Birks' parameter or energy
scale, enabling an in-situ investigation of the size of possible nonlinearity.  
Applying nonlinearity that is 20\% of the way from zero to the reference for
every simulated digitized charge
degrades the $\chi^2$ by one unit, with a correlated shift in
Birks' parameter. 
Thus at 25 MeV per plane (rightmost point in
Fig.~\ref{fig:birksprofile}) we do not have sensitivity to
nonlinearity effects with these data. 
The 20\% constraint is used
to add a component to the uncertainty for the Birks'
parameter measurement.  Because this constraint is correlated with
other aspects of this specific fit, the full upper bound of half the reference
suppression is kept as the uncertainty for all calorimetry analyses.

The Birks' measurement is consistent with other values for the Birks' quenching
parameter.  The parameter
value is expected to depend primarily on material formulation. 
A recent review of the properties of many materials including polystyrene
is available in \cite{Tretyak:2009sr} with references and one
additional later measurement \cite{Reichhart:2011gz}.   These
measurements are focused on heavily ionizing nuclear fragments and
alpha particles which are 
important in dark matter and double beta decay experiments as well as
nuclear fission studies.  The technique is conceptually
similar to using the end of a proton track but potentially more
sensitive due to the enhanced ionization and granularity of the data.  
The analysis of
\cite{Tretyak:2009sr} obtains a value of 0.0090 g / cm$^2$ MeV (with
no uncertainty given) for polystyrene based scintillator.  Using the
1.06 g/cm$^3$
density of polystyrene quoted in that analysis, this converts to 
0.085 mm/MeV.  This value and the Birks' parameter result above for our
scintillator formulation and density of 1.043 g/cm$^3$ are nearly identical.

\section{Proton calorimetry}

This test beam experiment is designed to constrain the uncertainty on
the single particle calorimetric response to protons and pions.   For
low-multiplicity neutrino events we reconstruct the hadron response
particle-by-particle using range, calorimetry, or a combination of the
two.  For high-multiplicity hadron systems from neutrino events, the total energy of the hadronic
recoil system (everything but the outgoing charged lepton) is
calorimetrically reconstructed.
When the hadron(s) interact in the detector, energy is spent unbinding
nucleons from nuclei and other energy goes to neutral particles.  
An estimate of this missing energy is used to correct the
observed response and obtain an unbiased estimator for the hadron
system.  In all cases, a major ingredient is the MC prediction for the
single particle response, the observed energy in the detector for a
given true energy, which is constrained with these data.  

The hadron event is reconstructed by summing the calibrated energy measured
in the scintillator.  The standard tracking algorithm is applied to each
event.  If a track segment is found, the 3D location of hits on the
track are known and used to make a correction
for attenuation in the scintillator strip to the point where the
particle passed.  For all hits not on tracks, the attenuation estimate is made to the
center of the strip.  Then a correction for the passive material
fraction for each plane is applied; 
a factor of 1.3 in the tracker, 2.1 in the ECAL, and 10.7 in the HCAL.  
Crosstalk is not included
when the muon equivalent technique is used to set the energy scale, 
but is measured as a byproduct of that calibration.
Because crosstalk is proportional to
the total of the energy deposits, the measured crosstalk fraction of 4.2\% is
subtracted from both data and MC.  

The activity recorded over the 150~ns 
integration time~\cite{Perdue:2012hg} is summed into the response, unlike the typical MINERvA neutrino
analysis which uses a window from -20~ns to +35~ns around the peak in the
cluster timing distribution.   Activity later than 150~ns from low energy neutrons and
decay electrons is not included.  The latter is predicted to amount to
a few percent of the available energy and appears in the detector
over several microseconds.

For the proton calorimetry analysis, the beamline-induced backgrounds
are reduced using additional selections.  For the lowest
proton energies, below 0.15~GeV for TE and 0.2~GeV for EH, the back half
of the detector is not included calorimetrically at all and is used as
a muon/pion veto by rejecting events with greater than 10 MeV of
activity.    Up to 0.3~GeV (TE) or 0.7~GeV (EH), backgrounds are reduced by using a 2
MeV threshold for activity in the last
four planes to veto background activity from the beam.  At the highest
energies, there is no background rejection.

\begin{figure}[ht]
\begin{center}
\includegraphics[width=6.0cm]{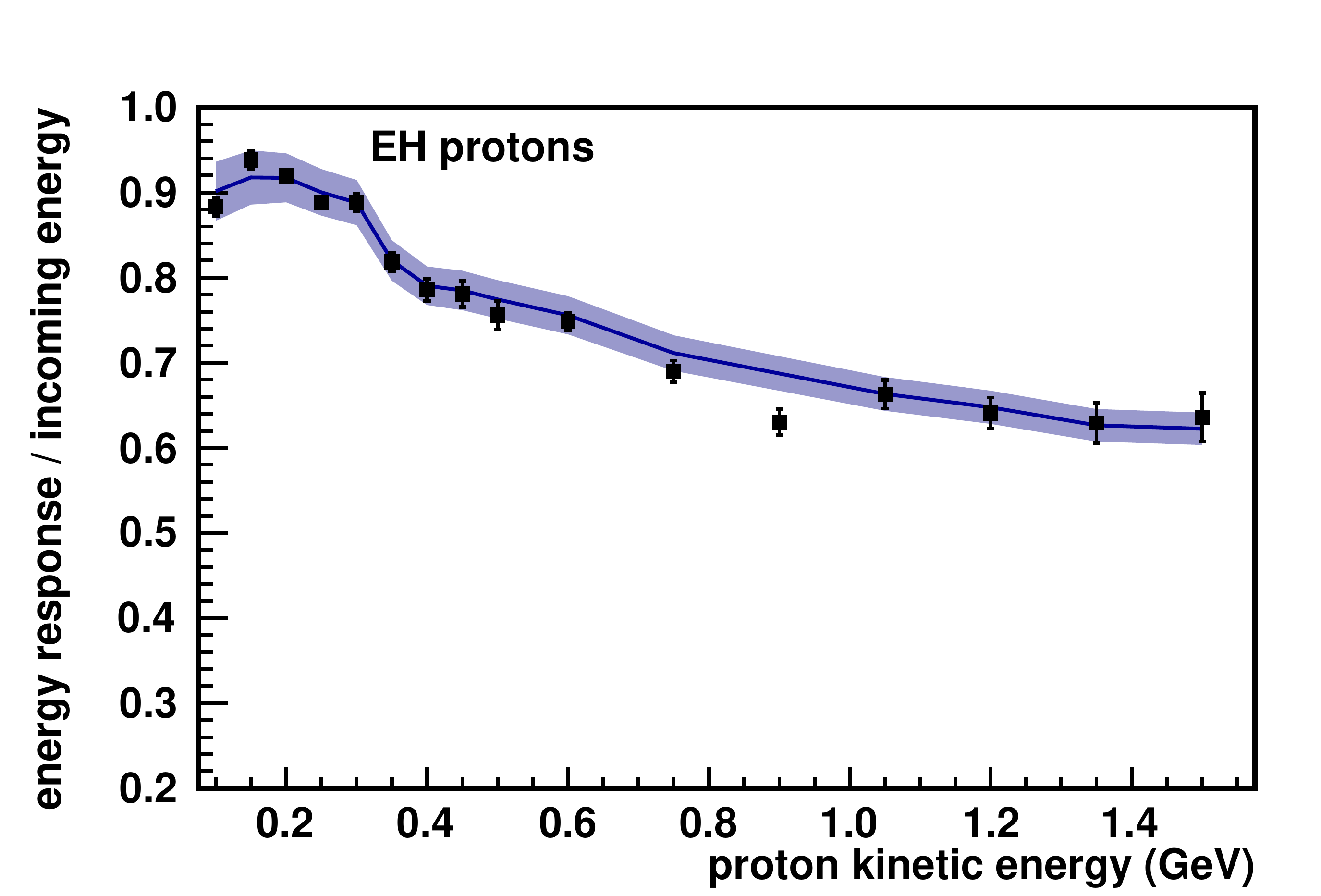}
\includegraphics[width=6.0cm]{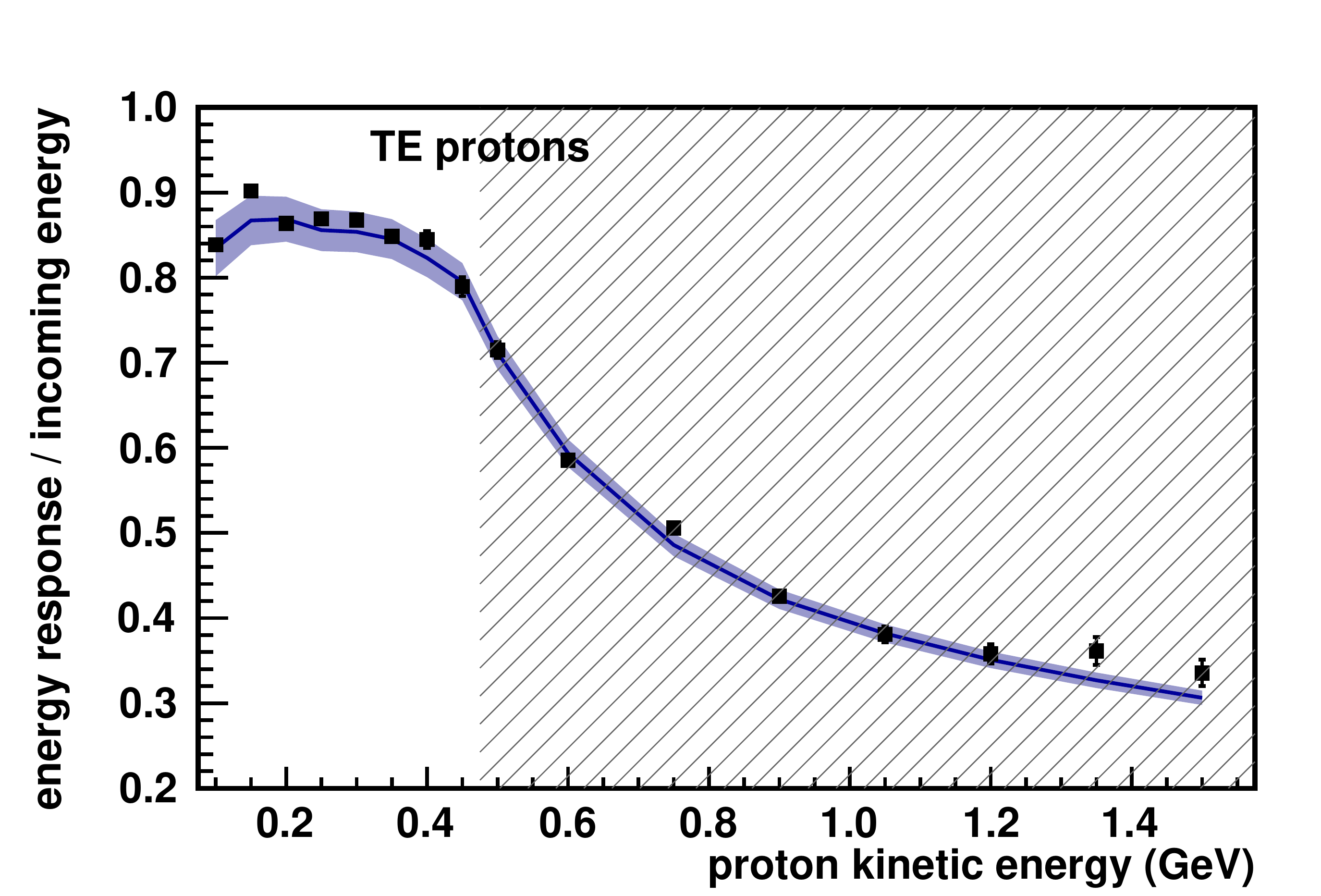}
\includegraphics[width=6.0cm]{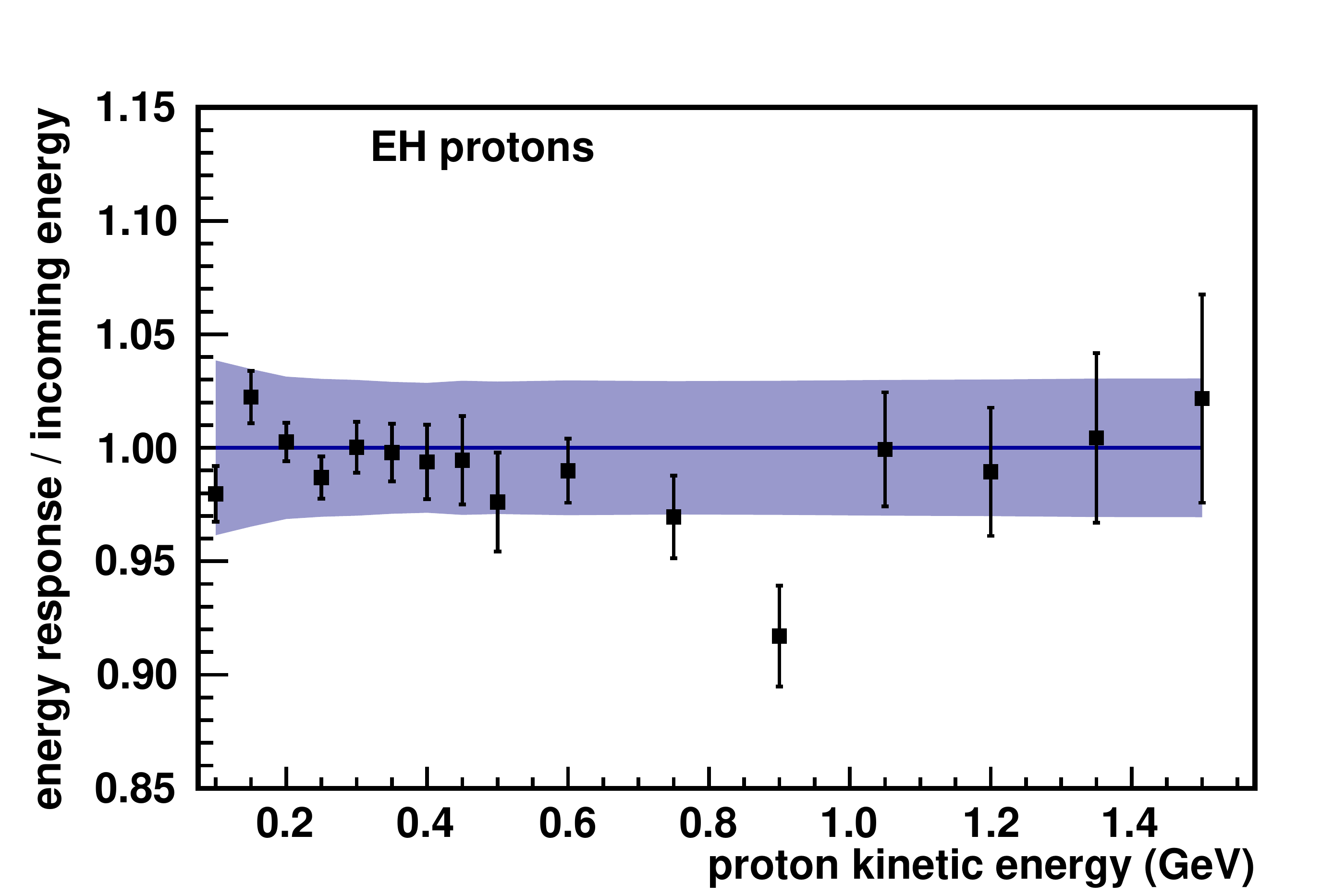}
\includegraphics[width=6.0cm]{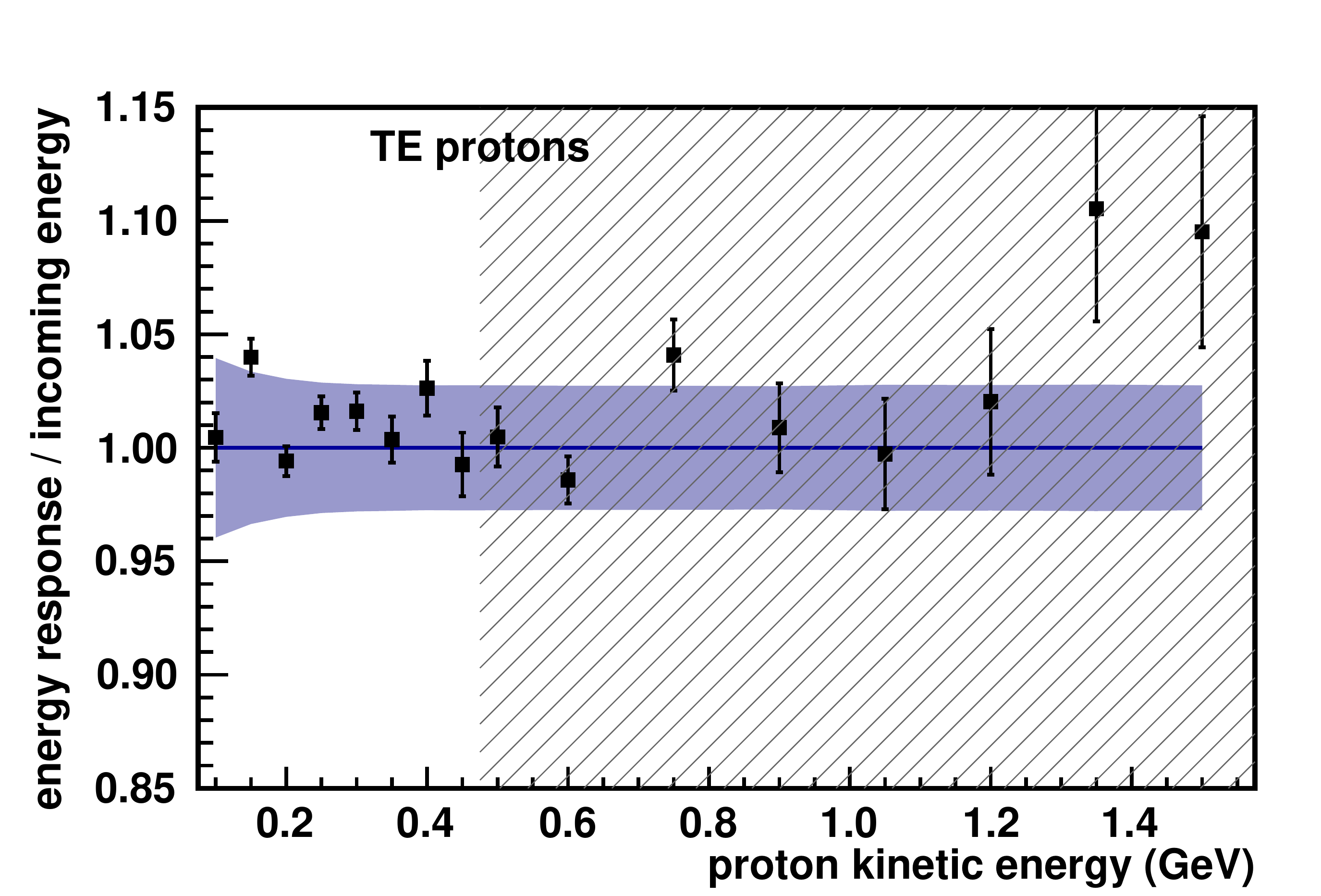}
\caption{ Proton response for EH (left) and
    TE (right).  The data points have statistical uncertainties, 
    the MC line has an error band with the systematic uncertainties described in
    Sec.~\ref{sec:systematics}.
    The bottom plots show the ratio data/MC. 
    The hatched TE region indicates energies where
    containment is so degraded that the measurement is not
    calorimetric.
\label{fig:protonresponse}}
\end{center}
\end{figure}

The resulting
corrected estimate for the energy is compared to the available energy,
which is just the kinetic energy for the proton.  The distribution of this fractional
response is the primary measurement and is calculated event-by-event.  
Then the events are binned by incident particle energy, and we compute
the mean and RMS for each bin.  The results for the mean are
plotted in Fig.~\ref{fig:protonresponse}. The error band on the MC represents
the total systematic uncertainty.  

The proton response has several features in this energy range.  At low
energy, the probability for a proton nuclear interaction is low.  As a
result, there is little missing energy, and also
the distribution of response is approximately Gaussian around its mean.  At
0.3~GeV, the protons begin to enter the HCAL in the EH detector and begin to produce $\Delta$
resonances when they interact in nuclei in both EH and TE configurations.  
Both lead to a drop in response, the former as the high dE/dx end of a
proton often happens in the steel, the latter because $\Delta$ production
generically leads to lower response through neutral final states and
energy lost to unbinding of additional nucleons.

The MC tracks the proton response well over the entire range.
The ratio data/MC for the mean response in
each energy bin is shown in Fig.~\ref{fig:protonresponse}.  The MC has negligible statistical uncertainty; the
systematic uncertainty on this ratio is shown as a band on the 
MC.
and described in detail in Sec.~\ref{sec:systematics}.
The data is shown with statistical uncertainties.
Despite a cut on time-of-flight applied to data and MC, 
there may be additional pion background at 0.15~GeV in the proton data because
those protons take 19~ns to travel the beamline.
These data points, and the data point at 0.9~GeV, stand out in the
figure of Fig.~\ref{fig:protonresponse}.
They correspond to no other special features of the
experimental setup, and have the character of fluctuations.

The response at low energy for the TE detector is partly correlated to the tuning of
Birks' parameter, because up to 0.25~GeV the two analyses use the same proton events.  However,
the strip response energy scale does not come from the free parameter
in the Birks' analysis, which would make this correlation even
greater.  
Instead, the muon equivalent unit calibration was redone
using the measured Birks' parameter to obtain the final strip energy
calibration.
Thus energy response offsets are correlated with the Birks'
parameter through its uncertainties, and less with the overlap of the data
events.

At higher energies for the TE configuration, hatched in Fig.~\ref{fig:protonresponse}, there is
a loss of containment of charged particles produced in the hadronic
interaction.
The calorimetric response no longer
represents the kind of result we expect for the larger MINERvA
detector.  Instead, these points demonstrate only that the MC is still
doing an adequate job describing the data.

\begin{figure}[ht]
\begin{center}
\includegraphics[width=6.0cm]{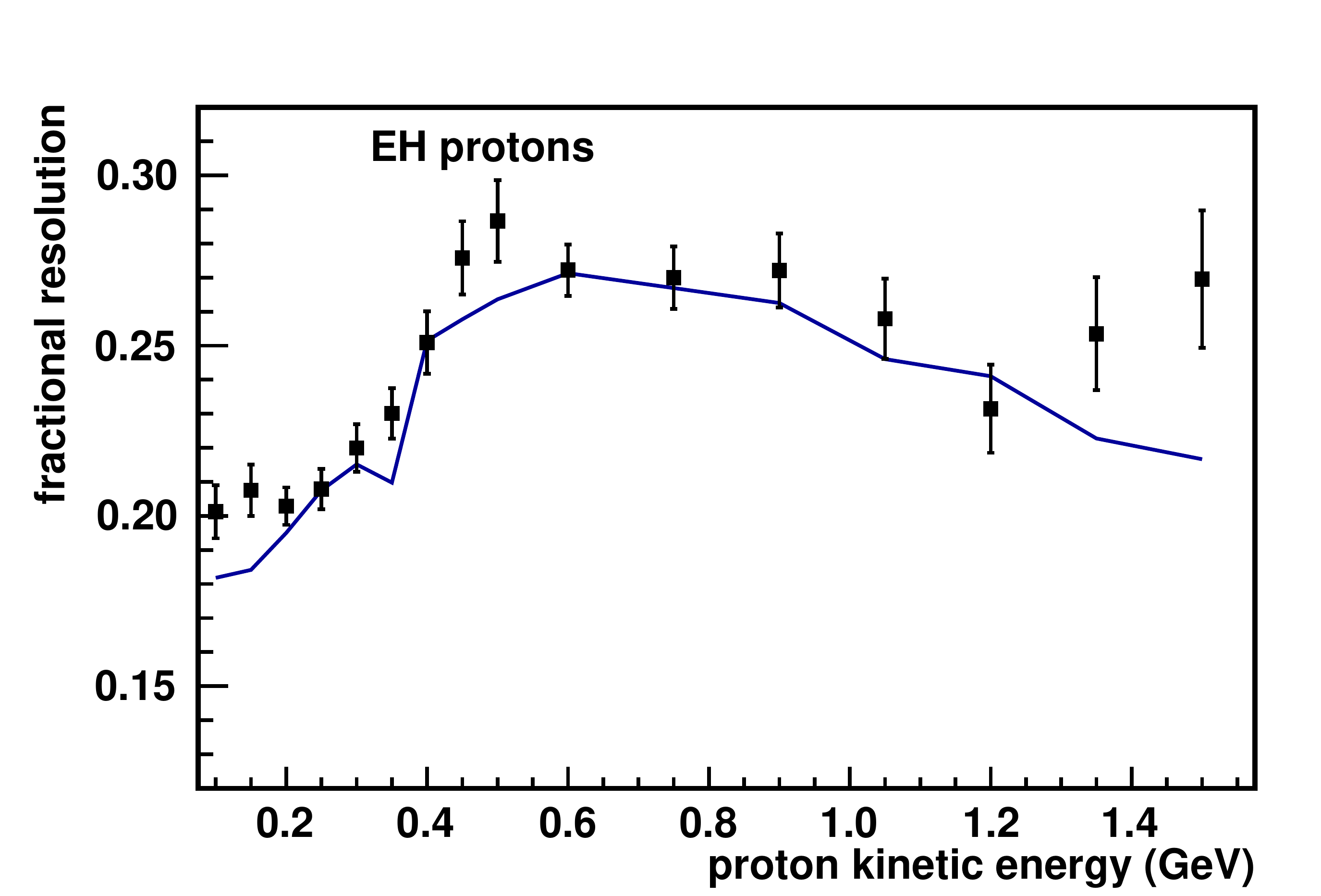}
\includegraphics[width=6.0cm]{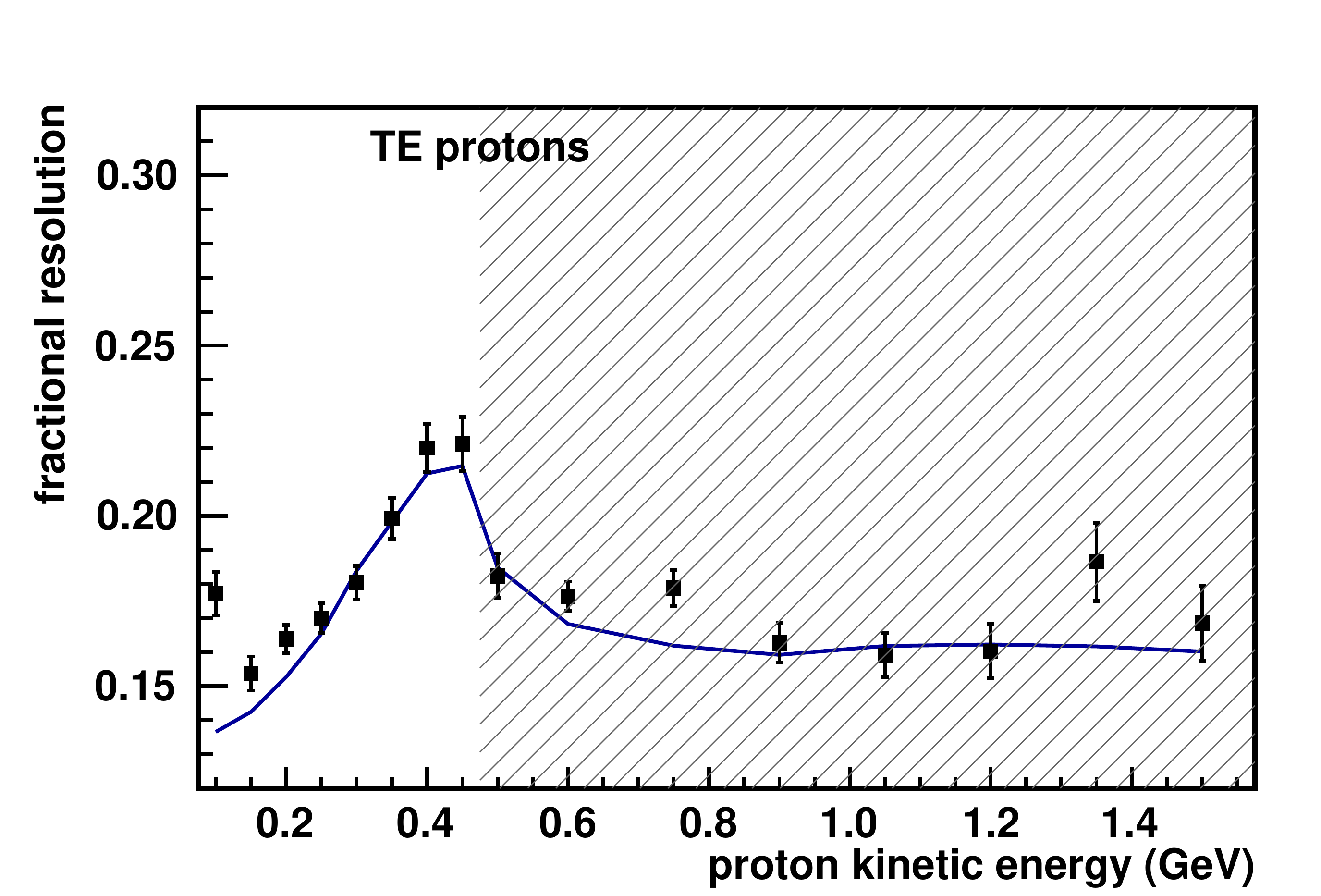}
\caption{ Fractional resolution, from the RMS of the proton calorimetric response for the EH configuration
  (left) and TE (right).   Only statistical uncertainties are shown on
  the data points, no uncertainties for the MC line.
  The hatched TE region indicates energies where
  containment is so degraded that the measurement is not
  calorimetric.
\label{fig:protonresolution}}
\end{center}
\end{figure}

In addition to the average response, it is important for MINERvA neutrino
analyses that event-by-event fluctuations in the response are well
simulated.  Many neutrino distributions are strongly peaked in
reconstructed energy or some other kinematic quantity, and an error in
resolution will flatten or sharpen the MC peak relative to the data,
causing a bias in unfolded distributions and fit parameters.
The basic shape of the distribution of response particle-by-particle
is well described, so it is adequate to use the RMS of the distribution
to quantify the trend and the agreement, as shown in
Fig.~\ref{fig:protonresolution}.  Only statistical uncertainties on the RMS are
shown, and no systematic uncertainty is quantitatively considered. 

At all energies, the MC response has a lower RMS, and more
so at low energy. 
The deviation can be taken to be a conservative uncertainty on the calorimetric
resolution.  A possible explanation for the degraded resolution in data
is the addition of beamline induced backgrounds which are not
simulated.  Such backgrounds are
not expected to have the same magnitude effect for higher energy
protons, the pion sample, or the
same origin as events in the MINERvA detector.

The resolution at 0.5~GeV is worse than at 0.3~GeV because of
two effects mentioned previously.
This is the region where proton interactions start to produce $\Delta$
resonances, which are also responsible for the decrease in response seen in the
upper left plot of Fig.~\ref{fig:protonresponse}.  Secondly, in the EH
configuration, this is the energy range where protons start to
reach the HCAL, and the high dE/dx endpoint of the proton is likely
in the iron.

Of special interest is the resolution for the lowest energy protons
which are contained in the tracker portion of the TE
detector configuration.  Such low energy protons are typically
found at the vertex of a neutrino interaction from
quasielastic and resonance production and include products of the
intranuclear rescattering process.  In the 0.05 to 0.2~GeV range, the
resolution is around 17\% and the distribution is nearly Gaussian.  
These protons are energetic enough to
travel through more than one plane but not energetic enough to excite
$\Delta$ resonances in the nucleus.  The largest contributions to the
resolution are from fluctuations at the end of the proton's range and (for data only)
from beam-induced background activity.
Above this energy, $\Delta$ production
becomes important, reducing the fraction of protons that stop at the
end of their range to about half the
total.  Also above 0.2~GeV, the distribution from which the RMS is computed picks up a
low-side tail whose shape is well-modeled by the MC.

\section{Systematic uncertainties for single particle response}
\label{sec:systematics}

The systematic uncertainties on the single particle response described
in this section also apply to the pion and electron measurements with only minor
differences. 
It is convenient to
present the systematic uncertainties together.  This completes the
discussion of the proton measurements, while providing information
which is helpful for interpreting the pion measurements
in Sec.~\ref{sec:pion}.
The significant sources of uncertainty are described in Table~\ref{tab:systematics}.

\begin{table}[ht]
\begin{center}
{\small
\begin{tabular}{lcccccc}
Source & TE p & EH p & EH $\pi^+$ & EH $\pi^-$   & EH e & TE e \\ \hline
Beam momentum & 1.9\% & 1.9\% & 1.0 to 2.0\% & 1.0 to 2.0\% & 1.0 & 1.0 \\
Beamline mass model & 0.7 & 0.7 & $<$0.2 & $<$0.2  & $<$0.2 & $<$0.2 \\
Birks' parameter   & 2.0 to 0.9 & 2.0 to 1.2 & 1.0 & 1.0 & 0.3 & 0.3 \\
Correlated late activity & 0.3 & 0.6 & 1.4 & 1.4 & $<$0.2 & $<$0.2 \\
Temperature stability & 1.0 & 1.0 & 1.0 & 1.0 & 1.0 & 1.0\\
Relative energy scale & 0.6 & 0.6 & 0.6 & 0.6 & 0.6 & 0.6 \\
PMT nonlinearity  & 0.7 & 0.7 & 0.9 & 0.9 & 0.4 & 0.2 \\
Event selection & $<$0.2 & $<$0.2 & 0.7 & 1.5 & 1.1 & 1.1 \\
Crosstalk & 0.7 & 0.9 & 0.5 & 0.5 & 0.5 & 0.5 \\ \hline
Statistical & $\sim$1.0 & $\sim$1.0 & $\sim$1.0 & $\sim$1.0 & 1.7 & 1.1 \\ \hline
Total & 3.3 to 2.7\% & 3.4 to 2.9\% & 2.6 to 3.4\% & 2.9 to 3.6\%  & 2.6\% & 2.3\% \\
                                                       
\end{tabular}
}
\caption{ Percent systematic uncertainties on the single particle
    fractional response for comparisons of the MC to data.  Additional uncertainties on
    the energy scale and absorber material apply 2.0\% equally to
    data and MC absolute response.  The total range represents the evolution with energy
    from 0.1 to 0.4~GeV for TE protons, 0.1 to 1.0~GeV for EH protons,
    and 0.4 to 2.0~GeV for both pion samples.  The statistical
    uncertainties for proton and pion response are shown in the
    figure for each data point, and are explicitly given in the table for both
    electron samples.
\label{tab:systematics}}
\end{center}
\end{table}

\subsection{Beam momentum}

This uncertainty is intrinsic to the design of the beam and the
estimate of the momentum of the incoming particle.  An uncertainty
here has the effect of shifting the denominator of the
fractional response.  The uncertainty in the incident particle
momentum
is derived from the wire chamber survey and the
measurement and simulation of the magnetic field.  
Because it is an uncertainty on the momentum, it
translates differently to uncertainties on the available particle energy
for protons and pions.   The lowest energy protons pick up an
additional 0.7\% uncertainty due to the energy loss in the material of the
beamline because they have higher ionization losses and those losses
are a larger fraction of the total.   With this and all other
uncertainties, any energy dependence is included the error bands in 
Fig.~\ref{fig:protonresponse} and Fig.~\ref{fig:pionresultssummary} and
the total even if not summarized in individual lines in Table~\ref{tab:systematics}.

\subsection{Birks' parameter}

Even after producing a best fit Birks' parameter in
Sec.~\ref{sec:birks}, the remaining improved
uncertainty is still one of the largest contributions to the accuracy
of the result.  Because low energy protons almost always have a high
dE/dx activity at the very end of the proton's range, and because that
activity is a larger fraction of the total energy for low energy
protons, that sample is most affected by this uncertainty.
The uncertainty in Birks' parameter
is treated as uncorrelated with the
energy scale and nonlinearity uncertainties.

\subsection{Correlated late activity}

Some uncertainties are revealed by varying event selection cuts.  
Proton response, and especially pion response,
changes when a cut is applied to remove events
when additional activity is reconstructed within 800~ns following the
triggered event.  The response in the MC, which has neither
beamline-induced backgrounds nor
PMT afterpulsing simulated, 
is higher because of the correlation with neutrons 
from the hadronic interaction(s), electrons from $\pi$ to $\mu$ to $e$
decay,  and other delayed activity.  
Activity beyond 150~ns from the trigger is not included in the calorimetric energy.
However, neutron activity preferentially follow pions with low fractional energy
response.  The response for the data is the opposite; it falls slightly and ends
about 1\% below the MC prediction.  Particles removed with
this cut in the data due to late, unrelated beamline activity should be
uncorrelated with the energy of the triggered event, and not bias the
mean response.   Instead, data particles with large shower activity and
possibly less missing energy generate more afterpulsing
and are more likely to have activity within the 800~ns after the event.
If the effect was primarily afterpulsing, the default selection is
optimal and this would not be a systematic uncertainty, but
an investigation did not confirm that hypothesis.  That the MC and
data disagree on how the response changes could be a
Geant4 modeling effect, which is what the experiment is designed to
measure.  However, we have not ruled out an experimental
effect, so this is included in the uncertainty.

\subsection{Temperature stability}

The response of the detector to cosmic ray muons for the data 
is calibrated against the measured
temperature in the experimental hall as a function of time.  This accounts
for the change over the course of the day and from day to day during
the run.  A correction is then applied to energy deposits in the beam
data, while the simulation has no
temperature dependence.  The uncertainty is estimated as the
difference between the responses of the high and low temperature
halves of the dataset, after the correction is applied.

\subsection{Relative energy scale}
\label{sec:relativeenergyscale}

The calibration procedure uses a comparison of simulated cosmic ray muons to
measured muons, so by construction the data/MC relative energy scale is
well constrained.  (The absolute energy scale is limited by our knowledge of the material
model for the scintillator planes and affects both data and MC.)  
The only significant contribution to this relative
uncertainty comes from observations of discrepancies between the TE
and EH data sets.
Within each subsample, there is no
discernable time-dependent trend in the energy response that can be extrapolated between these
two detector configurations.  The uncertainty listed here is taken
to be half the discrepancy seen in the muon calibrations between the
TE and EH data sets.

\subsection{PMT nonlinearity}

A nonlinearity reference curve is available from bench tests of these
photomultiplier tubes and is a suppression of response as a function of the total
measured charge.  Half the reference curve approximately
accounts for the translation from bench test conditions to detector
conditions with direct and reflected light.  The Birks' parameter measurement yields only an upper
bound for the magnitude of this effect, but that result is obscured by
correlations with other uncertainties.  We use half the reference
curve as the uncertainty here, applied to reduce the reconstructed
energy of the MC on a strip-by-strip basis.
Nonlinearity is a large effect for rare high activity strips, but
for hadronic tracks and showers at these low energies the overall
effect is modest.  This effect is one way because there is no PMT
nonlinearity in the simulation, so it serves only to move the
simulated energy lower.

\subsection{Event selection}
 
For protons, variations in the event selection do not produce
significant uncertainty, even near 0.15~GeV kinetic energy where the
19~ns pileup appears. 
There is an intrinsic electron and kaon background in the pion sample.
Variations in those selections yield a 0.7\% uncertainty for $\pi^+$
and twice the uncertainty for $\pi^-$.

\subsection{Crosstalk}

Optical and electronic crosstalk
in the cosmic muon calibration gives an average contribution of 4.2
$\pm$ 0.5\% to the energy in the detector, and the amount of crosstalk in the MC is
tuned to reproduce this.  Because the energy
calibration of the detector specifically does not include crosstalk,
the latter is subtracted from the total energy of each event.  The remaining 0.5\%
contributes directly to the calorimetric uncertainty between data
and MC.  Analysis of neutrino data also has crosstalk in the
simulation tuned to the data, but uses multiple techniques depending
on the analysis to deal
with crosstalk, including  thresholds, topological identification, and
subtraction.

\subsection{Absolute energy scale}

There are additional effects which apply equally to both data and MC
absolute energy scale and enhance the absolute uncertainty beyond to the relative
energy scale uncertainties.  The most important come from the
material model for the scintillator planes and also the lead and
iron absorber.  They affect both the calibration of the energy
deposits in the detector as well as how deep the hadronic activity propagates
into the detector.  They add an additional 2\% in quadrature to the
quantities in Table~\ref{tab:systematics} and the vertical axis in the
response figures for any situation where the absolute uncertainty is
needed.  The most important portion for calorimetry, from the
calibration of the energy scale, yields an uncertainty on the
calorimetric correction applied to both data and MC. 

\subsection{Geant4 step size}

The simulation is affected by a number of different Geant4 settings,
including some that are unrelated to the hadronic physics model.
 A setting of particular interest is the maximum step size allowed by
the Geant4 adaptive step size algorithm.
The baseline simulation uses essentially the default Geant4
settings, the same as used for the rest of the MINERvA experiment, so all the
calibrations and measurements are done with a consistent set of
parameters, and there is no uncertainty to assign.
Purposely making the
maximum step size 0.05 mm allows the adaptive algorithm to still
choose smaller steps near material boundaries but never larger steps.
This change results in a reduced MC response of 1\% for
pions and has no effect for 0.5~GeV/c electrons.  The effect is consistent with
causing an enhanced Birks' effect because then the simulation 
produces more highly quenched energy deposits; compare the opposite
study in Sec.~\ref{sec:birks} of 4\% enhancement in the last plane
with activity for a more coarse stepping.

\section{Pion calorimetry}
\label{sec:pion}

Two separate samples of pions were obtained by running the beam
magnets with different polarities.  The EH $\pi^+$ sample was
obtained concurrently with the proton sample while the $\pi^-$ sample
was from the data set taken the previous week.  After these data were
taken, the detector configuration was changed to the TE configuration,
but unlike for protons, containment in the TE is not adequate
for a pion calorimetry measurement.  Another difference is that the
lowest beam momenta available cause the lowest pion energy for this
analysis to be 0.35~GeV, just above the $\Delta$ production peak.  
The ECAL is less than one interaction length thick, but the HCAL is
more than one interaction length.  
Very few pions stop at the end of their range in the detector, but
many reach the HCAL before interacting.

The event selection and energy measurement proceed similar to the
proton case, including correcting the observed energy for passive material, 
crosstalk, and the last-four-plane veto.
The denominator for the fractional response for pions is taken to be
the total energy; some of the pion mass energy will become
reconstructed energy in the detector.  For pions there is a potential background at low
energy from electron contamination and at high energy from kaons (see
Fig.~\ref{fig:beamline})
which is neither simulated nor
subtracted.  
Variations in the selection process results in only small
changes to the response.

The background due to unrelated activity from the beam
has been estimated two ways. A measurement of activity 30~ns
earlier than the triggered particle gives one estimate.  For the lowest energy proton
sample, another estimate is made by measuring activity 
beyond plane 30 where there should
be negligible activity.  When extrapolating these estimates to the
whole detector and time of the event,
they both yield the same 4 MeV per event on average.  For
the mean response, this is simply subtracted from the total energy
before calculating the fractional response.  
At higher energy, the use of the last-four-plane veto leads to another
downward bias of about 1\% in the observed energy,
estimated using the MC, because real hadron interactions put energy
into those planes.  This bias is removed with a MC-based energy
dependent correction.  The pion analysis procedure is different than the use of stricter
cuts for protons but also leads to negligible uncertainty.

\begin{figure}[ht]
\begin{center}
\includegraphics[width=6.0cm]{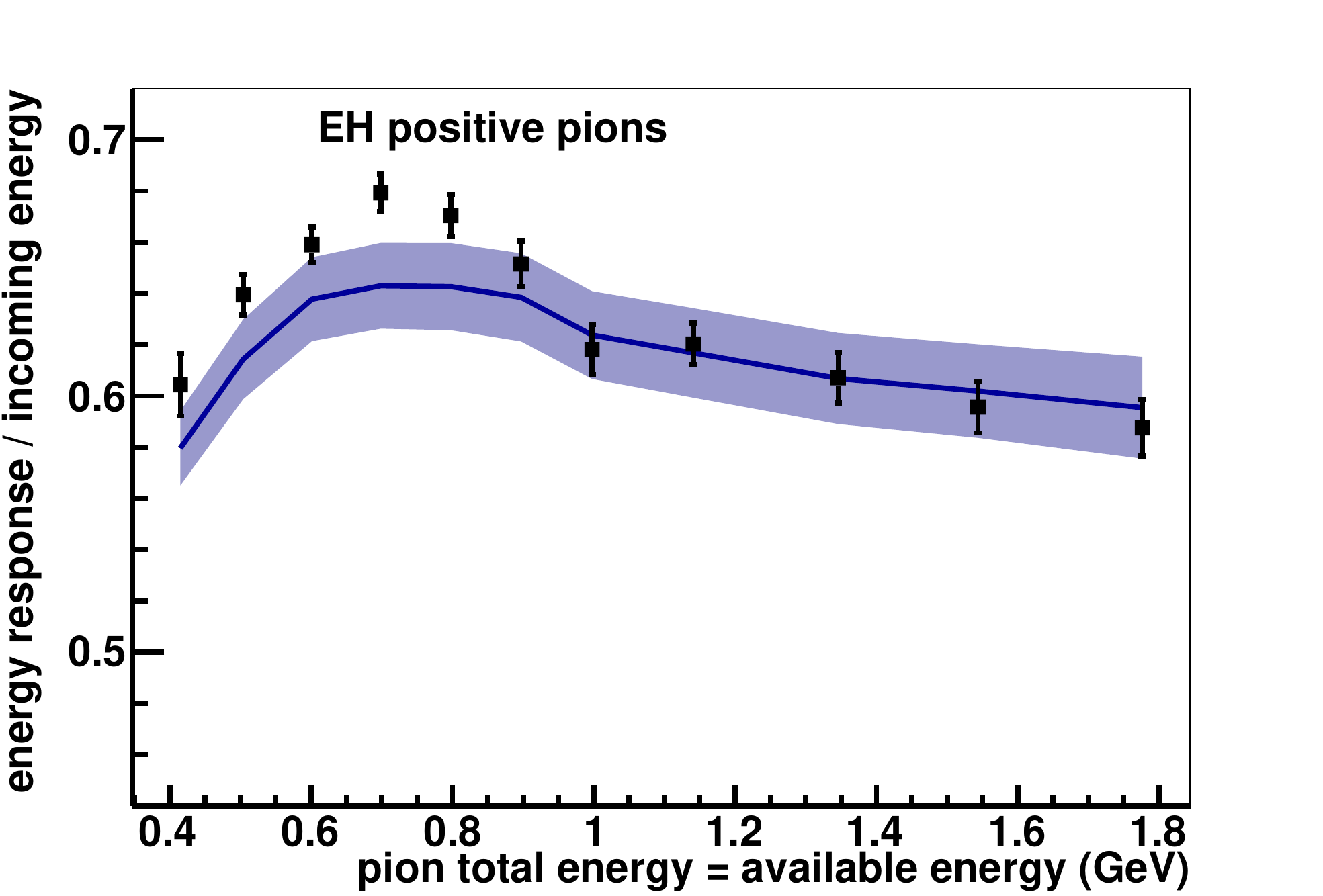}
\includegraphics[width=6.0cm]{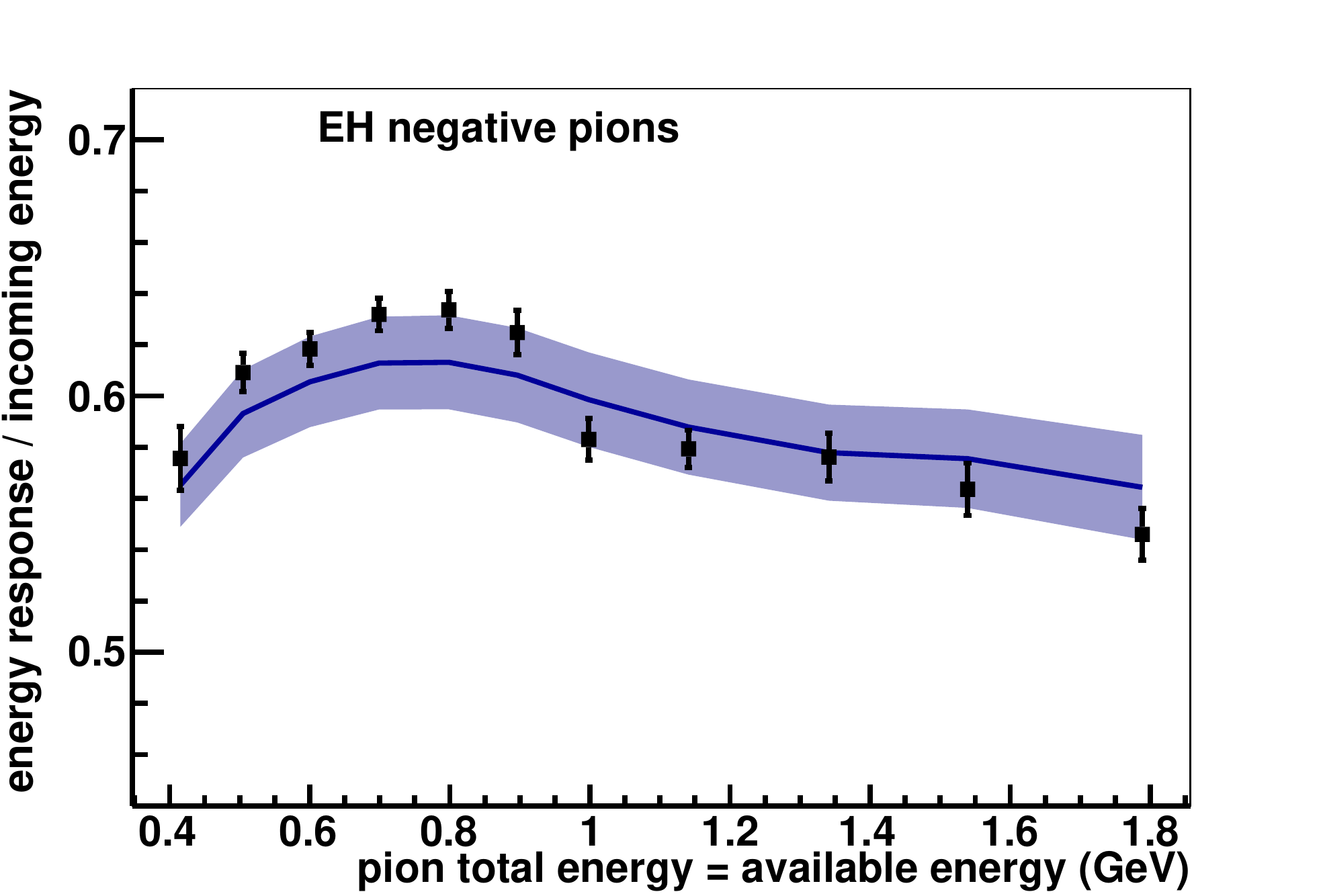}
\caption{Calorimetric response for positive (left) and
    negative (right) pions.  The errors on the data are statistical
    only, while the error band on the MC represents the systematic
    uncertainties associated with comparisons between data and MC.  
    A larger uncertainty of up to 4.2\% (not shown) applies to the absolute
    response scale for both data and MC.
\label{fig:pionresultssummary}}
\end{center}
\end{figure}

The MC describes the response to pions imperfectly, as
shown in Fig.~\ref{fig:pionresultssummary}.  The statistical
uncertainty on the data is shown but is negligible for the MC.  
Systematic uncertainties (with their energy
dependence) from Table~\ref{tab:systematics} are incorporated into the
MC error band.  The MC models
the single particle response to within 4\% averaging the points up to
1.0~GeV, and 3\% from there up to 2.0~GeV.

This level of agreement is adequate for MINERvA's
neutrino program, is used to assess the single-particle hadronic
response uncertainties for MINERvA analyses,
and no correction factor is needed.
However, the MC does not accurately model a change in
behavior that starts at 0.9~GeV, where there is a mild inflection
point in the MC predicted response.   The onset and
the magnitude of the discrepancy are the same for both $\pi^+$ and
$\pi^-$, equivalent to a 5\% decrease from low to high energy relative
to the MC.  The experimental systematic uncertainties permit some
shape distortion for higher energy relative to low,
about equally from the beamline uncertainties, 
species selection, and beamline-induced backgrounds.
When evaluated in quadrature, these could produce a $\pm$1.8\%
relative change over this energy range, less than half what is
observed.  None of these systematic uncertainties would naturally produce a change
over a short 0.2~GeV energy range near 0.9~GeV. 
If a future MINERvA
neutrino analysis is sensitive to this, we will need to parameterize
this effect instead of taking an overall uncertainty in the response.

In principle, these data are a test of not just our ability to model the
detector itself but also the ability to model the pion energy loss and reaction processes
such as inelastic, absorption, charge exchange, and elastic scattering.
We have investigated the sensitivity to model uncertainties
using the Bertini cascade model \cite{Heikkinen:2003sc} within Geant4, including consideration of pion cross
section data \cite{Ashery:1981tq,Allardyce:1973ce}.  However, calorimetry is more
sensitive to the total available energy than it is to differences in
the types of outcomes for the first particle-nucleon interaction in the hadronic shower.  Trial 30\%
modifications to the relative mix of outcomes have at most a 0.5\% effect
on calorimetry.
Instead, increasing the probability of pions to interact (either
elastically or inelastic with at least 10 MeV energy
transfer) before reaching the HCAL enhances the response.
By this definition of interaction, the mean free path in the ECAL is about 30 planes; lowering it by
20\% (increasing the Geant4 pion nucleus cross section) decreases the calorimetric response by 2\%.
An investigation of the trend reveals a correlation with the fraction of
events that have negligible energy in the HCAL:  the MC does not follow
the data and underestimates this fraction starting at 0.9~GeV.  
Such an underestimate is also a predicted effect of a too-high mean
free path, lowering it by 20\% increases shifts this fraction up 2.5\%.
Differences between models in Geant4 and reality in principle could be energy
dependent, so a tuned model could better describe the overall average
response or separately the anomalous trend with energy.

\bigskip

The ratio of detector response to positive pions over detector
response to negative pions cancels a number of common uncertainties and
the trends described in the preceding paragraphs.  The MC
predicts that $\pi^+$ yield a 4.8\% higher response than $\pi^-$.  The measured
ratio is 6.2\%, with no energy dependence for either data or MC.   
The statistical uncertainty in the ratio in data is only 0.5\% averaged over
all energies.  Another 0.6\% uncertainty in the data/MC relative energy scale comes
primarily from the unknown time or detector configuration dependent
effect described in Sec.~\ref{sec:relativeenergyscale}, which should
conservatively be applied to interpret this ratio.   There is no evidence for
either an intensity effect (the $\pi^+$ data was at higher intensity), or
an operational effect due to time or polarity in the beamline, nor a
temperature effect.  These uncertainties are judged by comparing two halves of each data
configuration further split along these operational parameters, though
these tests are themselves afflicted by 0.7\% statistical uncertainty.
This 6.2 - 4.8 = 1.4\% 
discrepancy is at two standard deviations, and it can be used as a
conservative uncertainty on the ratio, when applying it to neutrino analyses.

\begin{figure}[ht]
\begin{center}
\includegraphics[width=6.0cm]{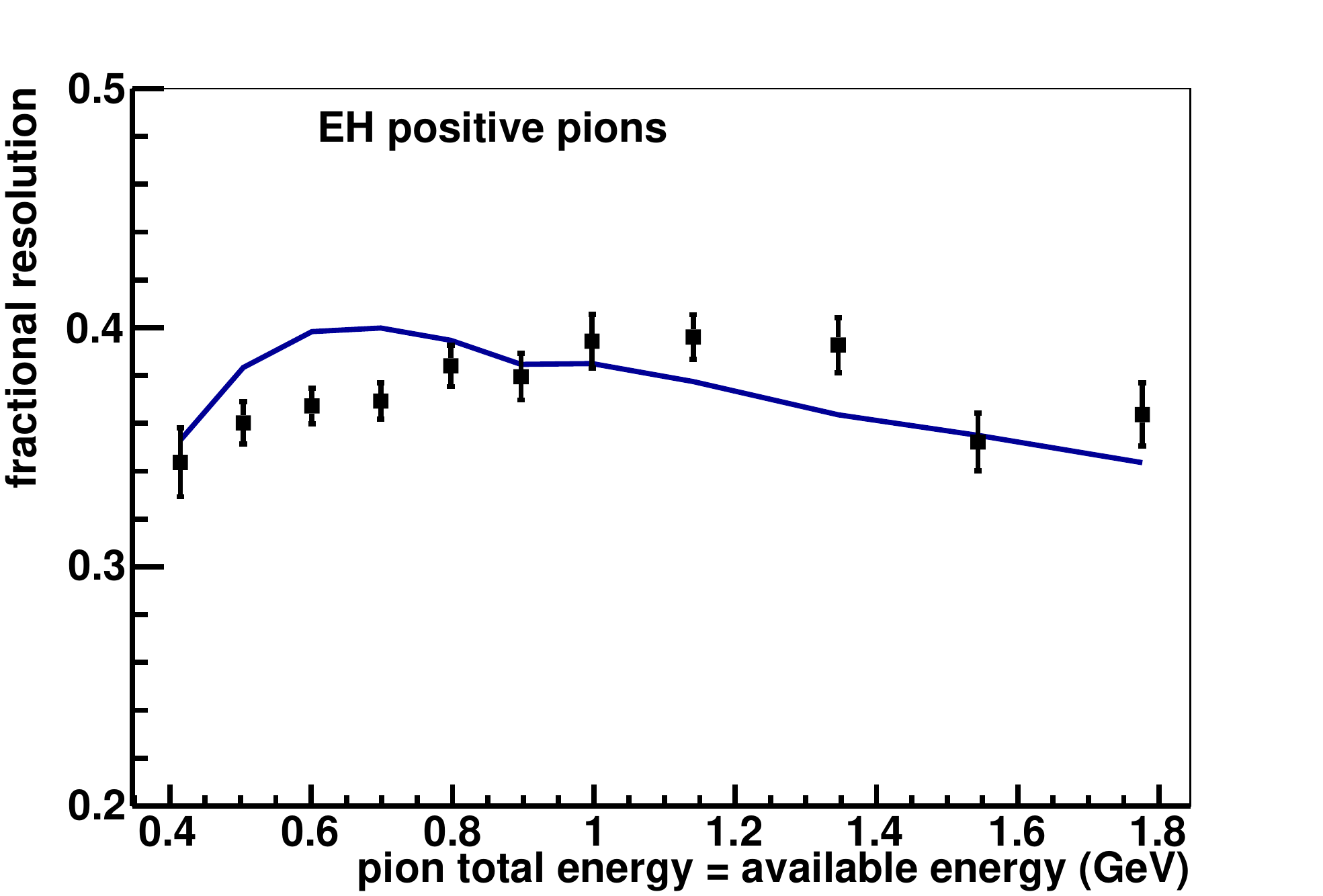}
\includegraphics[width=6.0cm]{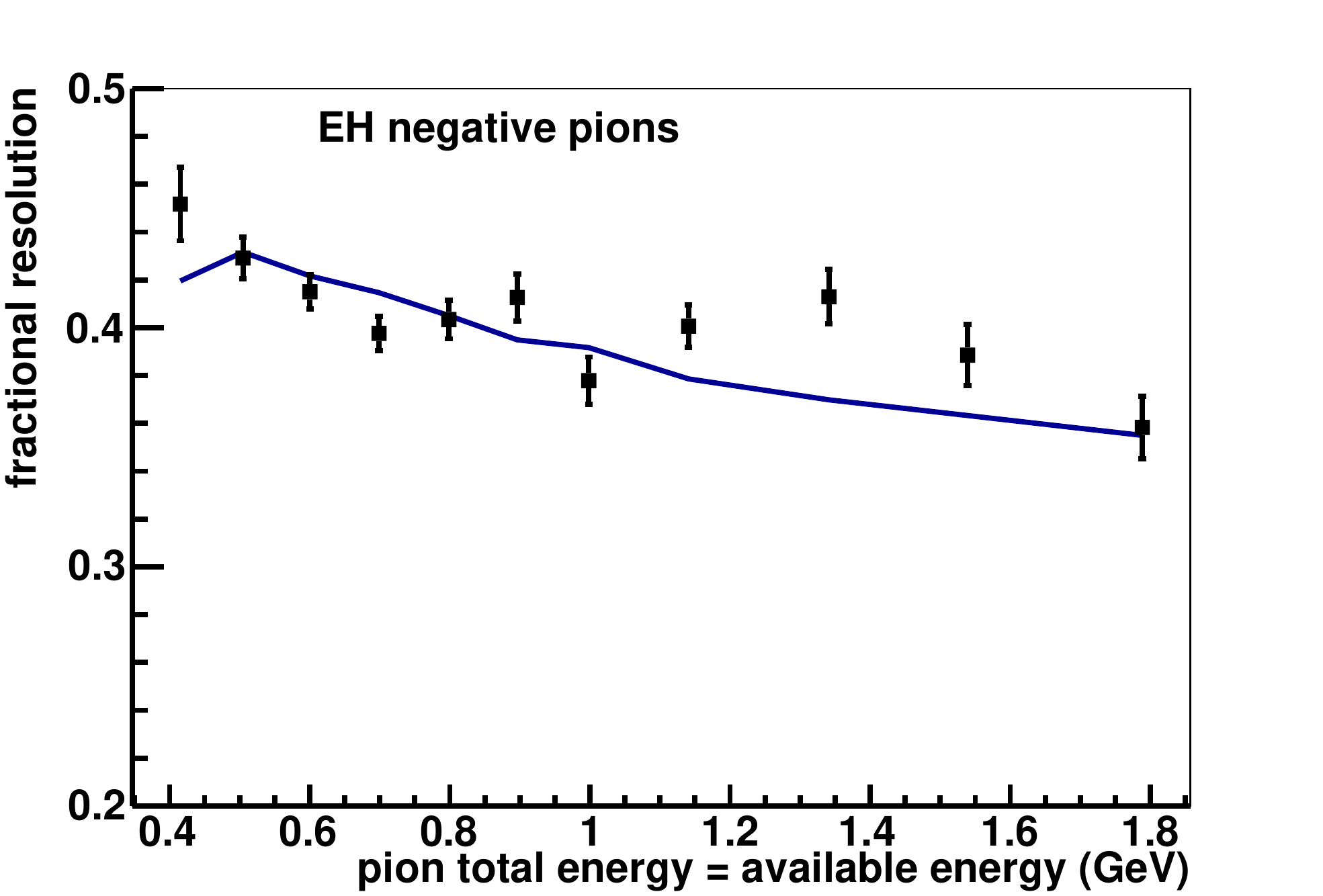}
\caption{Fractional resolution from the RMS of the calorimetric response for positive (left)
  and negative (right) pions.  The statistical error on the RMS is
  shown for the data points.  The predicted resolution from the MC is the
  line, and has no systematic uncertainties included.
\label{fig:resultsresolution}}
\end{center}
\end{figure}

As with the proton case, Fig.~\ref{fig:resultsresolution} shows
the resolution on the pion fractional response.  It is adequately modeled.
Beam-induced backgrounds are a much smaller fraction of the total
energy than for low energy protons, the $\Delta$ production peak
is at an energy below the lowest energy data, and a large fraction of the events
reach the HCAL,  so there is none of the structure seen in the proton case.

\section{Electron calorimetry}
\label{sec:electron}

The electron samples are limited to energies in a range from 0.4 to
0.5~GeV but are useful for studies with the ECAL portion of the detector.  
The production of electrons is intrinsically lower in energy and
fewer than pions and is predicted by an ab-initio simulation of the
beamline to be negligible for energies greater than 0.7~GeV.
Furthermore, the TOF resolution prevents good identification of the few that
are at higher energies.   In the EH detector configuration,
electrons in this energy range deposit 95\% or more of their energy in
the ECAL portion of the detector, and the response of the ECAL alone
can be measured.   The TE detector configuration is
similar: the electron propagates through the tracker but does not
shower extensively until the ECAL.

The electron sample is separated
from the pion sample using a combination of topological and
time-of-flight selections.  Events that resemble late-interacting pions
because they are tracked into the HCAL or because they have
a substantial fraction of energy in the back half of the detector are
rejected.  Further, the number of strips recording activity is
systematically more for electrons, and the variance in energy per
plane for EM showers is much higher than for interacting pions.
Using the MC, we estimate the efficiency for selecting
electrons (pions) to be 61\% (5\%) for the TE and 
73\% (8\%) for the EH configuration.  The pion and electron peaks separate in
time-of-flight by at least 0.7~ns at 0.5~GeV, easily separated given the 0.2~ns
resolution of the TOF measurement.   Extrapolating the pion
distribution just above the TOF cut into the selected electron region
in data yields an estimate of one pion background in 50 
electron events.  An eye-scan of the resulting events with the
web-based MINERvA event display \cite{Tagg:2011wk} yields 
one obvious background event which is removed, leaving 49 events total
in the EH sample.

The resulting sample is analyzed similarly as previously described for protons and
pions.   The data for electrons and positrons for
the EH configuration were combined into one sample; the MC is treated
the same way.
After correcting for passive material, crosstalk, and beamline-induced
background activity, the response ratio is
obtained for every event using the total electron energy as the denominator.
The electron fractional response is found to be 
0.763 $\pm$ 0.013 (statistical) in data
and 0.740 $\pm$ 0.002 (statistical) in MC.
There is an additional 2.0\% relative systematic uncertainty between
the data and MC, discussed in Table.~\ref{tab:systematics}, bringing the total uncertainty to 2.6\%.
Further adding uncertainties from the material assay brings this to
3.3\% absolute uncertainty.
The data response is 3\% higher than the MC predicts, a little more than the
total relative uncertainty.
The MC predicts a resolution of 11.5\%, which is an adequate
description of the low-statistics data.

The MC predicts the response in the TE configuration is 3\% higher
than the EH configuration
because most electrons ionize their way through the tracker before electromagnetic
showers develop in the ECAL.  The TE sample provides another 62 events,
with more positrons than electrons because of the running
conditions.   
Again a 3\% discrepancy response is seen in
these TE results, as with the EH results.  
The statistical uncertainty is smaller because of slightly better
statistics and
resolution.  The data/MC relative total uncertainty is 2.3\%, and the
absolute uncertainty on the response is 3.0\%.
The MC prediction of a 9.1\% resolution describes
the data well.

The electron sample analyses are subject to the same uncertainties as the proton and pion
results plus another 1.1\% uncertainty due to the additional
requirements to select electrons.
Table~\ref{tab:systematics} summarizes the
uncertainties in the final two columns.
Comparing the default MC to a variation with $\pm$1.2\% lead
density in the ECAL reveals only a
$\pm$0.15\% change in response for the TE configuration
and $\pm$0.3\% change for the EH sample.  Variations of the 
event selection contributes 1\% uncertainty to the response.  
The absolute energy scale uncertainty is the same 2\%
and the data vs. MC relative uncertainty is 0.7\%
from the material model effects and calibrations described previously.
Another 0.5\% comes from the crosstalk model.
Finally, the beam momentum uncertainty is 1\% at these energies.

\section{Calorimetry discussion}

In addition to the extensive studies of  high energy calorimetry
described by Richard Wigmans
in~\cite{Wigmans:2000vf}, there are several recent
test beam measurements using scintillator and
absorber sampling calorimeters and hadron simulations similar to MINERvA.  
Hadron calorimetry at energies below 2 GeV follows a process
where one hadron typically undergoes two, one, or zero inelastic
interactions, with a small number of outgoing charged particles.  
Unlike hadron calorimetry at higher energies,
the processes are not easily characterized by the statistical $\sqrt{E}$ and $\sqrt{N}$
effects.  

The MINOS neutrino experiment uses a detector made of scintillator and
inch-thick iron, very similar to the MINERvA HCAL.  Their test beam
exposures in the CERN T7 and T11 beamlines \cite{Adamson:2006xv} were
analyzed to produce electron \cite{Vahle:2004mp} and hadron
\cite{Kordosky:2004mn} calorimetry results, among other measurements
\cite{Cabrera:2009fi}.   They compared their data to
a {\small GEANT3} simulation and found several
discrepancies at the 3\% to 6\% level.  
However, our data are compared to a Geant4 simulation,
so interpretation relative to the present analysis is indirect.

The CALICE experiment has data from operating several kinds of
sampling calorimeters in beams at Fermilab and CERN.  They use
similar, Geant4 based hadronic and electromagnetic models, but
their data is mostly at higher energy.  The analysis of their data is
ongoing.  As of this writing, two publications
\cite{Adloff:2013kio,Adloff:2013jqa} can be compared with the
MINERvA test beam data.

Hadronic calorimetry is considered \cite{Adloff:2013kio} for data
taken with an iron-scintillator calorimeter.  They
find Geant4 models underestimate the measured response by 3\% at
8~GeV/c momentum, which is their lowest pion data available.  
This discrepancy is beyond the
edge of their 2\% uncertainty.  The 8~GeV/c data is also the only one
of the many model comparisons in their paper
where Geant4 is using the same Bertini cascade
model used in our simulation, 
shown in the lowest (blue) line in the lower left plot in their Fig.~6.  
Their data show a trend with energy such that the MC
overestimates the data above 20~GeV or so, but remain consistent within
their uncertainty estimates. 

In the later paper \cite{Adloff:2013jqa}, data from a tungsten segmented
calorimeter is compared to Geant4 models for electrons, pions,
and protons.  The $\pi^{+}$ response for the same Bertini cascade model (but from
Geant4 9.6.p2) describes
their mean response very well from 3 to 8~GeV.  The discrepancy is
less than 2\% while their uncertainty is around 3\%.  The
comparisons in this later paper include the same models 
and some of the same 
energies as in \cite{Adloff:2013kio}, but using data from a
different beam and an ECAL detector configuration.  
Agreement also follows
for proton data in the same range.  A similar result
is obtained for positrons, agreement above 2~GeV.  However, the simulation
underestimates the data by 2.5\% at 1~GeV,  just within one standard deviation agreement
for the lowest positron energy for which they have data.

Taken together, the MINERvA and CALICE data suggest that the Bertini
cascade model from recent (9.4p2 and later) Geant4 does a good job of
describing hadronic data at the 4\% level in an iron-scintillator
calorimeter through the combined range of energy.   CALICE indicates that the
electromagnetic cascade model applied to an ECAL style calorimeter
also does very well.
But the low energy data point
that is similar to MINERvA's suggests the MC underestimates the
response in both cases.

\section{Tracking efficiency}
\label{sec:tracking}


The proton sample in the TE detector configuration allows us to
measure the proton tracking efficiency, defined as the probability that a proton
will be reconstructed as a three-dimensional track object.   The
proton tracking efficiency, and that for
pions, is important for measurements of neutrino differential
cross sections with specific proton and pion final states.  

The sample is similar to the one used for the Birks' parameter
measurement where protons stop no later than plane 19, but without
the requirement that its depth be consistent with a proton at the end
of its range.  Another difference is that the sample is extended to protons whose
last activity is only as far as plane six.  
This analysis of tracking tests a combination of the standard
MINERvA ``long tracker'' which requires a minimum of eleven planes in
combination with either of two
variations of the ``short tracker'' which can
form tracks with as few as five planes of activity.  For this analysis, the MC sample is four times
the size of the data sample.  

The efficiency for long tracks is nearly perfect.  Specifically, the
sample of protons with kinetic energy less than 0.4~GeV whose
last energy deposit is 
between planes nine and nineteen (inclusive) are tracked with efficiency of 
99.2$^{+0.2}_{-0.3}$\% in data and 99.8$\pm$0.1\% in MC. 
For the data, this corresponds to tracking 1520 out of 1533 protons 
in the sample.
Around 60\% of protons stop a distance consistent with the end of
their range, and failing the tracking is highly
correlated with a proton experiencing an interaction.

Differences between the MC and data begin to appear for samples of
even shorter proton events.  For the
185 protons that appear to stop in plane eight, 178 of them were tracked,
which gives 96.2\% 
compared to the MC 97.7\% 
For 338 protons that appear to stop in planes six and seven only 308 are
tracked, 91.1\% 
compared to the MC 96.5\% 
These subsamples have a 70\% fraction with their stopping location at the
end of their expected range.  It is more likely in the data than the
MC that the subset of events with a short proton event
at the end of its expected range will not pass the tracking requirements.

\begin{table}
\begin{center}
\begin{tabular}{c|cc|cc}
proton & \multicolumn{2}{c|}{pion short tracker} & \multicolumn{2}{c}{quasielastic short tracker} \\
 depth & data & mc & data & mc \\ \hline
$\ge$ 9 planes & 99.2$^{+0.2}_{-0.3}$ \% & 99.8$\pm$0.1 \% &  99.5$^{+0.2}_{-0.2}$ \% &  99.9$\pm$0.1 \% \\
8 planes & 96.2$^{+1.2}_{-1.6}$ \% & 97.7$\pm$0.6 \%  &  96.8$^{+1.2}_{-1.6}$   \% & 98.3$\pm$0.5 \%  \\
6 and 7 planes & 91.1$^{+1.5}_{-1.6}$ \% & 96.5$\pm$0.5 \% &  93.5$^{+1.3}_{-1.4}$  \% & 98.1$\pm$0.4 \%  \\
\end{tabular}
\caption{ Summary of tracking efficiencies for the two configurations
  of the short-tracker combined with the same long-tracker algorithm.
\label{tab:trackingefficiencies}}
\end{center}
\end{table}

The above results for protons were obtained with a short tracker configured for a
neutrino pion production analysis \cite{Eberly:2014mra}.   A somewhat
different configuration optimized for a quasielastic proton analysis \cite{Walton:2014esl}
gives 1 to 2\% higher efficiency, successfully tracking an additional
6, 1, and 8 events in the data subsamples for the shortest, 8-plane,
and longest samples respectively, with a similar trend of better tracking
in the MC.   The efficiencies are summarized in
Table.~\ref{tab:trackingefficiencies}.

The main reason for the difference between the two tracking techniques
involves the choice of candidate clusters of activity to give to the
tracking algorithm.  The quasielastic proton algorithm is more permissive,
allowing clusters with more hits and more energy that would
be expected from a simply ionizing particle.  The pion algorithm excludes these
when deciding whether to form a track.  In the case of very short,
six-plane tracks, excluding one plane explains the difference between
the two algorithms.

Overall, the results suggest that tracking efficiency is adequately modeled
(within 1\%) for tracks greater than 9 planes, which makes it a
negligible uncertainty for neutrino analyses.  In contrast, we can use a
data-based correction of as much as 5\% to the efficiency for shorter
track lengths,
relative to the MC predicted efficiency.  In the MINERvA detector,
there is activity near the neutrino interaction point and wider range
of angles relative to the detector axis, which are not addressed by the
test beam sample.
Therefore, this efficiency correction should be on top of the MC
prediction for efficiency that considers other effects seen in real
neutrino interactions.

\section{Conclusion}

We have measured the performance of the tracking and calorimetry of the
MINERvA detector by exposing a scaled-down version of the
detector to a test beam
of low momentum protons, pions, and electrons from the Fermilab Test
Beam Facility.  These data provide a constraint on the Birks' law
saturation effect for our formulation of polystyrene based plastic
scintillator.  The calorimetric response to protons and pions within the range
of energies tested yields uncertainties of 4\% when the
single particle calorimetric response is used in neutrino analyses.
There are several effects that could be interpreted as two standard deviation fluctuations
relative to the systematic uncertainties, but overall the MC
describes the data and its resolutions well.  The electron
sample yields a similar uncertainty.  Tracking performance is well
modeled, and we have measured a small discrepancy between the
performance of tracking in the data and simulation.

\section*{Acknowledgements}

This work was supported by the Fermi National Accelerator
Laboratory under U.S. Department of Energy
Contract No. DE-AC02-07CH11359 which included the
MINERvA construction project. Construction of 
the test beam detector was granted by the United States National Science
Foundation under Grant No. PHY-0619727 and
by the University of Rochester. Support for participating
scientists was provided by NSF and DOE (USA)
by CAPES and CNPq (Brazil), by CoNaCyT (Mexico),
by CONICYT (Chile), by CONCYTEC, DGI-PUCP
and IDI/IGI-UNI (Peru), by Latin American Center for
Physics (CLAF), by the Swiss National Science Foundation,
and by RAS and the Russian Ministry of Education
and Science (Russia). 

These measurements were supported by the Fermilab Test Beam Facility staff,
and we particularly thank Doug Jensen, Erik Ramberg, Aria Soha 
for their support in design, installation, and operation of
the beam and experiment.  The finite element model analysis of the
magnetic field was done by Bob Wands at Fermilab.  Rob Napora assembled
the cosmic muon trigger.   We also acknowledge the late Bruno Gobbi for
his efforts refurbishing and early tests of the wire chambers, and overall guidance about instrumentation.

\section*{References}

\bibliography{testbeam-nim}

\end{document}